\documentclass[aps,prc,twocolumn,showpacs,preprintnumbers,amsmath,amssymb,nofootinbib]{revtex4-1}

\usepackage{sublabel}
\usepackage[]{graphicx}
\usepackage{epsf}
\usepackage{wrapfig}
\usepackage{epsfig}
\usepackage{wasysym}
\usepackage[usenames]{color}
\usepackage{pstricks}
\usepackage{amssymb, graphics}
\usepackage{indentfirst}
\usepackage{fancyhdr}
\usepackage{slashed}
\usepackage[latin1]{inputenc}
\usepackage[english]{babel}
\usepackage{amssymb}  
\usepackage{amsmath}  
\usepackage{latexsym} 
\usepackage{amsthm}   
\usepackage{subfigure}

\usepackage{color}

\begin{document}
\title{Comparative study of three-nucleon potentials in nuclear matter}
\author{Alessandro Lovato$^{1,2}$}
\author{Omar Benhar$^{2,3}$}
\author{Stefano Fantoni$^{4}$}
\author{Kevin E. Schmidt$^{5}$}

\affiliation
{
$^1$ SISSA, I-34014 Trieste, Italy \\
$^2$ INFN, Sezione di Roma. I-00185 Roma, Italy\\
$^3$ Dipartimento di Fisica, Universit\`a ``La Sapienza''. I-00185 Roma, Italy \\
$^4$ ANVUR, National Agency for the Evaluation of Universities and Research Institutes, Piazzzale Kennedy, 20. I-00144 Roma, Italy\\
$^5$ Department of Physics, Arizona State University, Tempe, AZ 85287, USA\\
}
\date{\today}
\begin{abstract}
A new generation of local three-body potentials providing an excellent description of the properties of light nuclei, 
as well as of the neutron-deuteron doublet scattering length, has been recently derived. We have performed a comparative analysis
of the equations of state of both pure neutron matter and symmetric nuclear matter obtained 
using these models of three-nucleon forces.
None of the considered potentials simultaneously explains the empirical equilibrium density and binding energy of symmetric nuclear matter. 
However, two of them provide reasonable values of the saturation density. 
The ambiguity   concerning the treatment of the contact term of the chiral inspired potentials is discussed.
\end{abstract}


\maketitle


\section{Introduction}

The definition of the potential describing three-nucleon interactions is a central issue of nuclear many-body theory. Three-nucleon forces (TNF) are long 
known to provide a sizable contribution to the energies of the ground and low-lying excited states of light-nuclei, and play a critical role in determining 
the equilibrium properties of symmetric nuclear matter. In addition, their effect is expected to become large, or even dominant, in high density neutron matter, the 
understanding of which is required for the theoretical description of compact stars. 

{\em Ab initio} nuclear many-body approaches are based on the premise that the dynamics can be modeled studying exactly solvable systems, 
having mass number $A \leq 3$. This is a most important feature since, due to the complexity of strong interactions and to the prohibitive
difficulties associated with the solution of the quantum mechanical many-body problem, theoretical calculations of nuclear observables generally 
involve a number of approximations. Hence, models of nuclear dynamics extracted from analyses of the properties of complex nuclei are 
plagued by the systematic uncertainty associated with the use of a specific approximation scheme.

Highly realistic two-nucleon potentials, either purely phenomenological \cite{lacombe_80,stocks_94,wiringa_95,machleidt_01} or  based on chiral perturbation theory (ChPT) \cite{entem_03,epelbaum_05}, have been obtained from accurate fits of the properties of the bound and scattering states of the two-nucleon system \cite{stocks_93,gabioud_79,vanderleun_82,ericson_83,rodning_90,simon_81,bishop_79}. Unfortunately, however, the extension to the case of the three-nucleon potential is not straightforward. 
Phenomenological models, such as the Urbana IX (UIX) potential, that reproduce the observed binding energy of $^3$H by construction, fail to explain the measured $nd$ doublet scattering length, $^2a_{nd}$ \cite{shoen_03}, as well as the proton analyzing power in $p$-$^3$He scattering, $A_y$ \cite{shimizu_95}.


In recent years, the scheme based on  ChPT has been extensively employed to obtain three-nucleon potential models \cite{epelbaum_02,{bernard_08}}. The main advantage of this approach is the possibility of treating the nucleon-nucleon (NN) potential and the TNF in a more consistent fashion, as the parameters $c_1$, $c_3$ and $c_4$, fixed by NN and $\pi N$ data, are also used in the definition of the TNF. In fact,  the next-to-next-to-leading-order (NNLO) three-nucleon interaction only involves two parameters, namely 
$c_D$ and $c_E$, that do not appear in the NN potential and have to be determined fitting low-energy three-nucleon (NNN) observables. Unfortunately, however, $\pi N$ and $NN$ 
data still leave some uncertainties on the $c_i$'s, that can not be completely determined by NNN observables.

A comprehensive comparison between purely phenomenological and {\em chiral inspired} TNF, which must necessarily involve the analysis of both pure neutron matter (PNM) and symmetric nuclear matter (SNM), is made difficult by the fact that chiral TNF are derived in momentum space, while many theoretical formalisms are based on the coordinate 
space representation.    

The local, coordinate space, form of the chiral NNLO three nucleon potential, hereafter referred to as  NNLOL, can be found in Ref. \cite{navratil_07}. 
However, establishing a connection between momentum and coordinate space representations involves some subtleties.  

The authors of Ref. \cite{epelbaum_02} have shown that the NNLO (momentum space) three body potential obtained from the chiral Lagrangian, when operating on a 
antisymmetric wave function, gives rise to contributions that are not all independent of one another. To obtain a local potential in coordinate space one
has to regularize using the momenta transferred among the nucleons. This regularization procedure makes all the terms of the chiral 
potential independent, so that, in principle, all of them have to be taken into account. The potential would otherwise be somewhat inconsistent, as it becomes apparent 
in nuclear matter calculations, which involve larger momenta.

A comparative study of different three-nucleon local interactions (Urbana UIX (UIX), chiral inspired revision of Tucson-Melbourne (TM$^\prime$) and chiral NNLOL three body potential), used in conjunction with the local  Argonne $v_{18}$ NN potential, has been recently performed \cite{kievsky_10}. 
The authors of Ref.~\cite{kievsky_10} used the hyperspherical harmonics formalism to compute the binding energies of $^3$H and $^4$He, as well as the $nd$ doublet scattering length, and found that the three body potentials do not
simultaneously reproduce these quantities. Selecting different sets of parameters for each TNF they were able to obtain results compatible with experimental data, although a unique parametrization for each potential has not been found.  This problem is a consequence of the fact that the three low-energy observables considered are not enough to completely fix the set of parameters entering the definition of the potentials.

The work described in this paper is aimed at testing the different parametrization of the potentials in nuclear matter.  In the case of SNM, a {\em realistic} Equation of State (EoS) 
is constrained by the available empirical information on saturation density, $\rho_0$, binding energy per nucleon at equilibrium, $E_0$, and compressibility, $K$. 
Furthermore, the recent observation of a neutron star of about two solar masses \cite{demorest_10} puts a constraint on the stiffness of the EoS of beta-stable matter, closely 
related to that of PNM. 

Nuclear matter calculations are carried out using a variety of many-body approaches. 
The scheme referred to as FHNC/SOC, based on correlated basis functions and the cluster expansion technique, has been first used to perform accurate nuclear matter calculations
with realistic three body potentials in Ref.~ \cite{carlson_83}. This analysis included early versions of both the Urbana (UIV, UV) and Tucson Melbourne (TM) three 
body interactions with the set of parameters reported in  Ref. \cite{coon_81}. 
The results indicate that the UV model, the only one featuring a phenomenological repulsive term, provides a reasonable nuclear matter saturation density, while 
the UIV and TM potentials fail to predict saturation. In addition, none of the considered models yields reasonable values of the SNM binding energy and compressibility. 

The findings of Ref.~\cite{carlson_83} are similar to those obtained in Ref.~\cite{wiringa_88}, whose authors took into account additional diagrams of the 
cluster expansion and used the UVII model. The state-of-the-art variational calculations discussed in Ref.~\cite{akmal_98}, carried out using the 
Argonne $v_{18}$ \cite{wiringa_95} and UIX \cite{pudliner_95} potentials, also sizably underbinds SNM. While the authors of Ref.~\cite{akmal_98} ascribed 
this discrepancy to deficiencies of the variational wave function, the analysis of Refs.~\cite{gandolfi_07,lovato_11} suggest that this problem can be largely 
due to the uncertainties associated with the description of three-nucleon interactions, whose contribution turns out to be significant.

Momentum space chiral three-body interaction have been also employed in nuclear matter \cite{bogner_05,hebeler_11,hebeler_10}. In these studies,   
the NNNLO chiral two-body potential has been evolved to low momentum interaction $V_{low\,k}$, suitable for standard perturbation theory in 
the Fermi gas basis. The results, showing that the TNF is essential to obtain saturation and realistic equilibrium properties of SNM \cite{bogner_05,hebeler_11}, 
exhibit a sizable cutoff dependence. At densities around the saturation point this effect is $\sim 4\,\text{MeV}$. 
In addition, different values of the constants $c_i$ lead to different Equations of State for SNM \cite{hebeler_11} and PNM \cite{hebeler_10}. 

The main features of the chiral inspired TNF are briefly reviewed in Section \ref{TBF}, while in Section \ref{many-body} we analyze the coordinate space 
form of the TNF derived in Ref.~\cite{kievsky_10} and discuss several issues related to the calculation of their contributions in nuclear matter, both within the 
FHNC/SOC and the Auxiliary Field Diffusion Monte Carlo (AFDMC) approaches.  The numerical results, including the EoS of PNM and SNM are discussed 
in Section \ref{results}. Finally, in Section \ref{conclusions} we summarize our findings and state the conclusions.  

\section{Chiral inspired models of three nucleon forces}
\label{TBF}
\begin{figure}[!ht]
\includegraphics[width=9.0cm,angle=0]{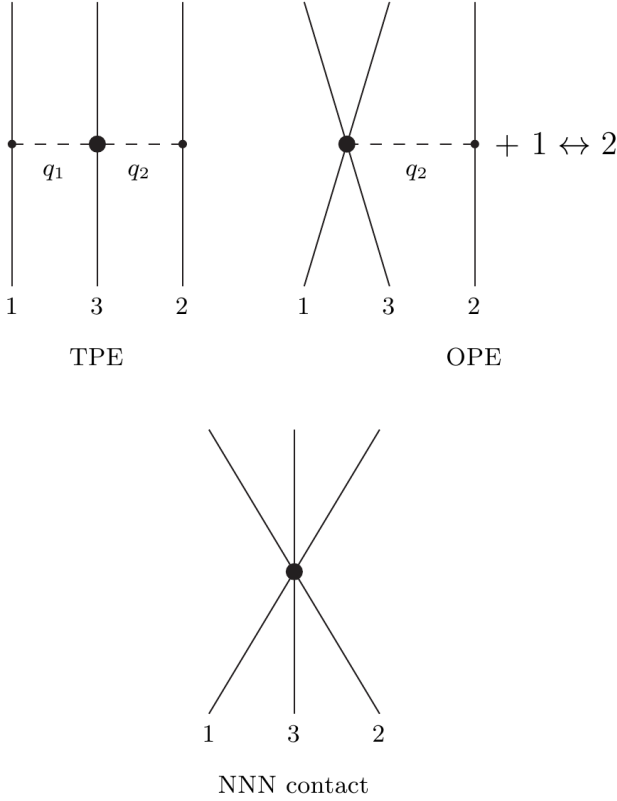}
\caption{TPE, OPE and NNN contact interactions of the chiral three-body force at NNLO.}
\label{fig:NNLO_diag}
\end{figure}

In a chiral theory without $\Delta$ degrees of freedom, the first nonvanishing three-nuclon interactions appear at NNLO in the Weinberg power counting scheme \cite{weinberg_90,weinberg_91}. The interaction  is described by three different physical mechanisms, corresponding to three different topologies of Feynman diagrams, drawn in Fig. \ref{fig:NNLO_diag} 
\cite{epelbaum_02}. The first two diagrams correspond to two-pion exchange (TPE) and one-pion exchange (OPE) with the pion emitted (or absorbed) by a contact NN interaction. The third diagram represents a contact three-nucleon interaction.

The full expression for the TNF is obtained by summing all possible permutations of the three nucleons. For this kind of potential, it turns out that there are only three independent cyclic permutations, i.e.
\begin{align}
V^{\chi}(1,2,3)&=V^{\chi}(1:2,3)+V^{\chi}(2:1,3)\nonumber \\
               &+V^{\chi}(3:1,2)
\end{align}
The Feynman diagrams of Fig. \ref{fig:NNLO_diag} refer to the permutation $(3:12)$, that can be written as
\begin{align}
V^{\chi}(3:12)&=c_1V_1(3:12)+c_3V_3(3:12)+c_4V_4(3:12)\nonumber\\
             &+c_DV_D(3:12)+c_EV_E(3:12)\, .
\end{align}
The first three terms $V_1$, $V_3$ and $V_4$ come from the TPE diagram and are related to $\pi N$ scattering. 
In particular, $V_1$ describes the $S$-wave contribution, while $V_3$ and $V_4$ are associated with the $P$-wave. The other terms, $V_D$ and $V_E$, are the OPE and 
contact contributions, respectively. 
Their momentum space expressions are \cite{epelbaum_02} 
\begin{align}
\tilde{V}_1(3:12)&=-V_0m_{\pi}^2\,\tau_{12}\frac{(\boldsymbol{\sigma}_1\cdot\mathbf{q}_1)}{(q_{1}^2+m_{\pi}^2)}
\frac{(\boldsymbol{\sigma}_2\cdot\mathbf{q}_2)}{(q_{2}^2+m_{\pi}^2)}\nonumber\\
\tilde{V}_3(3:12)&=\frac{V_0}{2}\,\tau_{12}\frac{(\boldsymbol{\sigma}_1\cdot\mathbf{q}_1)}{(q_{1}^2+m_{\pi}^2)}
\frac{(\boldsymbol{\sigma}_2\cdot\mathbf{q}_2)}{(q_{2}^2+m_{\pi}^2)}\mathbf{q}_1\cdot\mathbf{q}_2\nonumber\\
\tilde{V}_4(3:12)&=\frac{V_0}{4}\,\boldsymbol{\tau}_3\cdot(\boldsymbol{\tau}_1\times\boldsymbol{\tau}_2)
\frac{(\boldsymbol{\sigma}_1\cdot\mathbf{q}_1)}{(q_{1}^2+m_{\pi}^2)}
\frac{(\boldsymbol{\sigma}_2\cdot\mathbf{q}_2)}{(q_{2}^2+m_{\pi}^2)}\nonumber \\
&\qquad\boldsymbol{\sigma}_3\cdot(\mathbf{q}_1\times\mathbf{q}_2)\nonumber\\
\tilde{V}_D(3:12)&=-V_0^{D}\tau_{12}\Big[\frac{(\boldsymbol{\sigma}_2\cdot\mathbf{q}_2)}{(q_{2}^2+m_{\pi}^2)}(\boldsymbol{\sigma}_1\cdot\mathbf{q}_2)\nonumber\\
&\quad+\frac{(\boldsymbol{\sigma}_1\cdot\mathbf{q}_1)}{(q_{1}^2+m_{\pi}^2)}(\boldsymbol{\sigma}_2\cdot\mathbf{q}_1)\Big]\nonumber\\
\tilde{V}_E(3:12)&=V_0^{E}\tau_{12}\, ,
\end{align}
where $\boldsymbol{\tau}_i$ and $\boldsymbol{\tau}_i$ are the Pauli matrices describing the spin and the isospin of particle $i$. With $\sigma_{ij}$ and $\tau_{ij}$ we denote the scalar product $\boldsymbol{\sigma}_i\cdot \boldsymbol{\sigma}_j$ and $\boldsymbol{\tau}_i\cdot \boldsymbol{\tau}_j$, respectively.
The strengths of the TPE, OPE and contact terms, $V_0$, $V_{0}^D$ and $V_{0}^E$ are given by
\begin{equation}
V_0=\Big(\frac{g_A}{F_{\pi}^2}\Big)^2\qquad V_{0}^D=\frac{g_A}{8 F_{\pi}^4 \Lambda_\chi}\qquad V_{0}^E=\frac{1}{F_{\pi}^4 \Lambda_\chi} 
\end{equation}
where $g_A=1.29$ is the axial-vector coupling constant, $F_\pi=92.4\,\text{MeV}$ is the weak pion decay constant and $\Lambda_\chi$ is
the chiral symmetry-breaking scale, of the order of the $\rho$ meson mass. 

The low energy constants (LEC) $c_1$, $c_2$ and $c_3$ also appear in the sub-leading two-pion exchange term of the chiral NN potential and
are fixed by $\pi N$ \cite{fettes_98,buttiker_00} and/or $NN$ \cite{entem_03} data.
The parameters $c_D$ and $c_E$ are specific to the three-nucleon interaction and have to be fixed using NNN low energy observables, such as the $^3$H binding energy and 
the $nd$ doublet scattering length $^2a_{nd}$ \cite{epelbaum_02}.

The many-body methods employed in our work, namely FHNC/SOC and AFDMC, require a local expression of the three-body potential in coordinate space, 
that can be obtained performing  the Fourier transform \cite{navratil_07} 
\begin{align}
V^{\chi}(3:12)&=\int \frac{d^3q_1}{(2\pi)^3}\frac{d^3q_2}{(2\pi)^3} \tilde{V}^{\chi}(3:12)\times\nonumber \\
&\qquad F_{\Lambda}(q_{1}^2)F_{\Lambda}(q_{2}^2) {\rm e}^{i{\bf q}_1 \cdot {\bf r}_{13}}{\rm e}^{i{\bf q}_2 \cdot {\bf r}_{23}} \ ,
\end{align}
where the cutoff functions $F_{\Lambda}$, defined as
\begin{equation}
F_{\Lambda}(q_{i}^2)=\text{exp}\Big(-\frac{q_{i}^4}{\Lambda^4}\Big) \ ,
\label{eq:chi_cut}
\end{equation}
can depend on the momenta transferred among the nucleons, $q_i$, only. This feature has important consequences for the OPE and contact terms, that 
will be discussed at a later stage. 

The cutoff $\Lambda$ in the previous equation, while not being required to be the same as  $\Lambda_\chi$,  is of the same order of magnitude. 
Choosing the fourth power of the momentum in Eq.~(\ref{eq:chi_cut}) is therefore convenient, as the regulator generates powers of $q/\Lambda$ which 
are beyond NNLO in the chiral expansion. 

The Fourier transform can be readily computed, and provides the following coordinate-space representation of the chiral three-body potential:
\begin{align}
V_1(3:12)&=W_0\,\tau_{12}(\boldsymbol{\sigma}_1\cdot\vec{r}_{13})(\boldsymbol{\sigma}_2\cdot\vec{r}_{23})y(r_{13})y(r_{23})\nonumber\\
V_3(3:12)&=W_0\,\tau_{12}[\sigma_{12}y(r_{13})y(r_{23})\nonumber \\
&+(\boldsymbol{\sigma}_{1}\cdot\vec{r}_{23})(\boldsymbol{\sigma}_{2}\cdot\vec{r}_{23})t(r_{23})y(r_{13})\nonumber\\
&+(\sigma_{1}\cdot\vec{r}_{13})(\sigma_{2}\cdot\vec{r}_{13})t(r_{13})y(r_{23})\nonumber\\
&+(\vec{r}_{13}\cdot\vec{r}_{23})(\sigma_{1}\cdot\vec{r}_{13})(\sigma_{2}\cdot\vec{r}_{23})t(r_{13})t(r_{23})]\nonumber\\
V_4(3:12)&=W_0\,(\boldsymbol{\tau}_3\cdot\boldsymbol{\tau}_1\times\boldsymbol{\tau}_2)
[(\boldsymbol{\sigma}_{3}\cdot\boldsymbol{\sigma}_{2}\times\boldsymbol{\sigma}_{1})y(r_{13})y(r_{23})\nonumber\\
&+(\boldsymbol{\sigma}_{3}\cdot\vec{r}_{23}\times\boldsymbol{\sigma}_1)(\boldsymbol{\sigma}_2\cdot\vec{r}_{23})
t(r_{23})y(r_{13})\nonumber\\
&+(\boldsymbol{\sigma}_2\cdot\vec{r}_{13}\times\boldsymbol{\sigma}_3)(\boldsymbol{\sigma}_1\cdot\vec{r}_{13})
t(r_{13})y(r_{23})\nonumber \\
&+(\boldsymbol{\sigma}_3\cdot\vec{r}_{23}\times\vec{r}_{13})(\sigma_{1}\cdot\vec{r}_{13})(\sigma_{2}\cdot\vec{r}_{23})
t(r_{13})t(r_{23})]\nonumber\\
V_D(3:12)&=W_{0}^D\tau_{12}[\sigma_{12}y(r_{23})z_{0}(r_{13})\nonumber\\
&+(\boldsymbol{\sigma}_1\cdot\vec{r}_{23})(\boldsymbol{\sigma}_2\cdot\vec{r}_{23})t(r_{23})z_{0}(r_{13})\nonumber\\
&+\sigma_{12}y(r_{13})z_{0}(r_{23})\nonumber\\
&+(\boldsymbol{\sigma}_2\cdot\vec{r}_{13})(\boldsymbol{\sigma}_1\cdot\vec{r}_{13})t(r_{13})z_{0}(r_{23})]\nonumber\\
V_E(3:12)&=W_{0}^E\tau_{12}z_{0}(r_{13})z_{0}(r_{23})\ , 
\label{eq:2pi_3body_conf}
\end{align}
where $W_0$, $W_{0}^D$ and $W_{0}^E$ are obtained multiplying the corresponding $V_0$, $V_{0}^D$ and $V_{0}^E$ by a factor $m_{\pi}^6/(4\pi)^2$.
The radial functions appearing in the above equations are defined as 
\begin{align}
y(r)&=\frac{z_{1}'(r)}{r}\nonumber\\
t(r)&=\frac{1}{r^2}\Big(z_{1}''(r)-\frac{z_{1}'(r)}{r}\Big)=\frac{1}{r}y'(r)
\label{eq:def_y_t}
\end{align}
while $z_n$, proportional to $Z_n$ introduced in Ref.~\cite{coon_81}, is given by
\begin{align}
z_n(r)&=\frac{4\pi}{m_{\pi}^3}\int\frac{d^3q}{(2\pi)^3}\frac{F_\Lambda(q^2)}{(q^2+m_{\pi}^2)^n}e^{i\mathbf{q}\cdot\mathbf{r}}\nonumber\\
&=\frac{2}{\pi m_{\pi}^3}\int dq q^2\frac{F_{\Lambda}(q^2)}{(q^2+m_{\pi}^2)^n}j_0(qr)\, ,
\label{eq:def_zn}
\end{align}
with $j_0(x)=\sin(x)/x$. 
Note that, due to the form of the cutoff function of Eq.~(\ref{eq:chi_cut}), the radial functions are not known in analytic form, and must be 
obtained from a numerical integration.

Recently, the authors of Ref.\cite{kievsky_10} have studied the low energy NNN observables using the hyperspherical harmonics formalism and 
a nuclear hamiltonian including the NNLOL potential and the Argonne $v_{18}$ \cite{wiringa_95} two-body interaction. 
This mixed approach requires a fit of all the LEC appearing in the chiral  three-body interaction, not $c_D$ and $c_E$ only.  The best fit parameters for the $^3$H and $^4$He binding energies and for the $nd$ scattering length, $^2a_{nd}$, are listed in Table \ref{tab:chi_parameters}. For all the different parametrizations, denoted 
by NNLOL$_i$, $c_1$ and $\Lambda_\chi$ have been fixed to their original values $0.00081\, \text{MeV}^{-1}$  and $700\,\text{MeV}$, respectively \cite{epelbaum_02}. 
The momentum cutoff of Eq.~(\ref{eq:chi_cut}) has been set to $500\,\text{MeV}$.

\begin{table}[h!]
\begin{center}
\caption{Parameters of the NNLOL interactions of Ref.~\cite{kievsky_10}. \label{tab:chi_parameters}}
\vspace{0.3cm}
\begin{tabular}{c c c c c c} 
\hline 
 Potential &  $c_3\,(\text{MeV}^{-1})$  &  $c_4\,(\text{MeV}^{-1})$  &  $c_D$  &  $c_E$ \\ 
\hline
NNLOL$_1$ & -0.00448 & -0.001963 & -0.5 & 0.100  \\ 
NNLOL$_2$ & -0.00448 & -0.002044 & -1.0 & 0.000  \\ 
NNLOL$_3$ & -0.00480 & -0.002017 & -1.0 &-0.030 \\
NNLOL$_4$ & -0.00544 & -0.004860 & -2.0 &-0.500 \\
\hline
\end{tabular} 
\vspace{0.1cm}
\end{center}
\end{table}

As noticed in Ref.~\cite{friar_99}, despite the different underlying physical mechanisms, both TM and UIX three-nucleon interactions can be written as a sum of terms of the same form as those appearing in Eq.~(\ref{eq:2pi_3body_conf}). The differences among NNLOL, TM and UIX lie in the constants and in the radial functions. 

The TM$^\prime$ potential only involves the $V_1$, $V_3$ and $V_4$ contributions \cite{coon_01}. The cutoff function for this potential is not the same as in 
Eq.~(\ref{eq:chi_cut}), but 
\begin{equation}
F_{\Lambda}(q^2)=\Big(\frac{\Lambda^2-m_{\pi}^2}{\Lambda^2+q^2}\Big)^2\,.
\label{eq:TM_cut}
\end{equation}
The above form allows for the analytical integration of Eq.~(\ref{eq:def_zn}), yielding the radial functions
\begin{align}
y(r)&=\frac{e^{-r\Lambda}}{2m_{\pi}^3 r^3}\Big[2-m_{\pi}^2r^2-2(1+m_\pi r)e^{r(\Lambda-m_\pi)}\nonumber\\
&\quad+r\Lambda(2+r\Lambda)\Big]\nonumber\\
t(r)&=\frac{e^{-r\Lambda}}{2m_{\pi}^3 r^5}\Big[-6+2(3+3m_\pi r+m_{\pi}^2r^2)e^{r(\Lambda-m_\pi)}\nonumber \\
&\quad+m_{\pi}^2r^2(1+r\Lambda)-r\Lambda[6+r\Lambda(3+r\Lambda)]\Big]\, .
\end{align}

The TM$^\prime$ potential corresponds to the following choice of the strength constants (compare to Eq.~(\ref{eq:2pi_3body_conf}))
\begin{align} 
W_{0}=\Big(\frac{g m_\pi}{8\pi m_{N}}\Big)^2m_{\pi}^4
\end{align}
and
\begin{equation}
c_1 = \frac{a}{m_{\pi}^2} \quad,\quad c_3=2b\quad,\quad c_4=-4d \ ,
\end{equation}
$a$, $b$ and $c$ being the parameters entering the definition of the TM$^\prime$ potential \cite{coon_01}. 
The authors of Ref.\cite{kievsky_10} have determined the parameters of the TM$^\prime$ potential fitting the same set of low energy NNN observables
employed for the NNLOL potential.
In order to get a better description of the experimental data, they introduced a repulsive three-nucleon contact term, similar to the 
chiral $V_E$, but with $\tau_{12}$ omitted
\begin{equation}
V_{E}(3:12)=W_{0}^{E}z_0(r_{13})z_0(r_{23})\, ,
\end{equation}
where
\begin{equation}
W_{0}^{E}=\Big(\frac{g m_\pi}{8\pi m_{N}}\Big)^2\frac{9 m_{\pi}^2}{\Lambda_\chi}\, .
\label{eq:V0ETM}
\end{equation}
The corresponding radial function can be computed analytically from Eq.~(\ref{eq:def_zn})
\begin{equation}
z_{0}(r)=\frac{e^{-r\Lambda}}{8\pi\Lambda}(m_{\pi}^2-\Lambda^2)^2\, .
\end{equation}

As in the original paper \cite{coon_81}, in Ref.~\cite{kievsky_10} the value of the pion-nucleon coupling constant is set to $g^2=179.7\,\text{MeV}$,  the pion mass is  $m_\pi=139.6\,\text{MeV}$ and the nucleon mass is defined through the ratio $m_N/m_\pi=6.726$. The symmetry breaking scale $\Lambda_\chi$ of Eq.~(\ref{eq:V0ETM}) has the same value,  $700\,\text{MeV}$,  
used for the NNLOL potential.

The parameters of the TM$^\prime$ potentials, TM$^\prime_i$, that according to Ref.\cite{kievsky_10} reproduce the binding energies of $^3$H and  $^4$He and $^2a_{nd}$, are listed in Table \ref{tab:tm_parameters}. It turns out that $V_1$, gives a very small contribution to the low energy NNN observables. Therefore, the parameter $a$ has been 
kept to its original value $-0.87\,m_{\pi}^{-1}$. 

\begin{table}[h!]
\begin{center}
\caption{Parameters of the TM$^\prime$ potential reproducing low energy the NNN experimental data with $a=-0.87\,m_{\pi}^{-1}$
 \cite{kievsky_10}. \label{tab:tm_parameters}}
\vspace{0.3cm}
\begin{tabular}{c c c c c c} 
\hline 
 Potential & $b(m_{\pi}^{-3})$ & $d(m_{\pi}^{-3})$ & $c_E$ & $\Lambda(m_\pi)$\\ 
\hline
TM$^\prime_1$  & -8.256 & -4.690 & 1.0 & 4.0 \\ 
TM$^\prime_2$  & -3.870 & -3.375 & 1.6 & 4.8 \\ 
TM$^\prime_3$  & -2.064 & -2.279 & 2.0 & 5.6 \\
\hline
\end{tabular} 
\vspace{0.1cm}
\end{center}
\end{table}

The Fujita Miyazawa term \cite{fujita_57} of the UIX potential \cite{pudliner_95}, $V^{2\pi}$, describing the process 
whereby two pions are exchanged among nucleons and a $\Delta$ resonance is excited in the intermediate state, is conveniently written as
\begin{align}
\hat{V}^{2\pi}(3:12)&=A_{2\pi}\{\hat{X}_{13},\hat{X}_{23}\}\{\tau_{13},\tau_{23}\}\nonumber\\
&+C_{2\pi}[\hat{X}_{13},\hat{X}_{23}][\tau_{13},\tau_{23}]\,,
\end{align}
where
\begin{equation}
\hat{X}_{ij}=Y(m_\pi r)\sigma_{ij}+T(m_\pi r)S_{ij}\, ,
\end{equation}
and 
\begin{equation}
S_{ij}=3(\hat{r}_{ij}\cdot \boldsymbol{\sigma}_i)(\hat{r}_{ij}\cdot \boldsymbol{\sigma}_j)-\sigma_{ij}\, 
\end{equation}
is the tensor operator.
The radial functions associated with the spin and tensor components read
\begin{align}
Y(x)&=\frac{e^{-x}}{x}\xi_Y(x)\\
T(x)&=\Big(1+\frac{3}{x}+\frac{3}{x^2}\Big)Y(x)\xi_T(x) 
\label{eq:YT}
\end{align}
and the $\xi(x)$ are short-range cutoff functions defined as
\begin{equation}
\xi_{Y}(x)=\xi_{T}(x)=1-e^{-cx^2}\, .
\end{equation}
In the original derivation of the UIX potential the ratio $C_{2\pi}/A_{2\pi}$ was fixed to $1/4$ and the cutoff parameter was $c=2.1$ fm$^{-2}$, the same value as in the cutoff functions of the one-pion exchange term of the Argonne $v_{18}$ two-body potential. 

It can be shown that the anticommutator and commutator terms correspond to $V_3$ and $V_4$ of Eq.~(\ref{eq:2pi_3body_conf}), provided the following relations between the constants
\begin{align}
bW_0&=4A_{2\pi}\nonumber\\
dW_0&=4C_{2\pi}\, 
\label{eq:const_rel}
\end{align}
and the radial functions
\begin{equation}
\left\{
\begin{array}{rl}
Y(r)&=y(r)+\frac{r^2}{3}t(r)\\
T(r)&=\frac{r^2}{3}t(r)
\end{array} \right.
\label{eq:rad_gen_UIX}
\end{equation}
are satisfied.

The repulsive term of the UIX potential
\begin{equation}
V^R(3:12)=U_0 T^2(m_\pi r_{13})T^2(m_\pi r_{23})\, ,
\label{eq:VR}
\end{equation}
is equivalent to the $V_E$ term appearing in the TM$^\prime$ potential and (aside from the $\tau_{12}$ factor) in the NNLOL chiral potential if the following relations 
hold
\begin{equation}
T^2(m_\pi r)=z_0(r)\quad,\quad U_0=c_E W_{0}^E\, .
\label{eq:E_equiv}
\end{equation}

The UIX potential was not designed to reproduce low energy NNN observables only. While the parameter $A_{2\pi}$ was obtained from the fit of the observed binding energy of $^3$H, the strength $U_0$, was indeed adjusted to reproduce the empirical saturation density of SNM, $\rho_0=0.16\,\text{fm}^{-3}$.

In Ref. \cite{kievsky_10} it has been found that the original parametrization of the UIX potential underestimates $^2a_{nd}$ and slightly overbinds  of $^4$He.

\section{Three Nucleon Potentials in nuclear matter}
\label{many-body}
The investigation of uniform nuclear matter may shed light on both the nature and the parametrization of the TNF, although the 
quantitative description of this system can not be achieved within a mere generalization of the approaches developed for light nuclei. 
In this Section, we analyze the structure of the contact term of the NNLOL potential of  Ref. \cite{kievsky_10} and discuss the calculation of the TNF contribution to 
nuclear matter energy.

\subsection{NNLOL contact term issue}\
\label{sec:cont_issue}

While the  NNLOL chiral interactions provide a fully consistent description of the binding energies of $^3$H and $^4$He, as well as of the scattering length $^2a_{nd}$, some ambiguities emerge when these interactions are used to calculate the nuclear matter EoS. 

For our purposes, it is convenient to rewrite the NNLOL chiral contact term of Eq, (\ref{eq:2pi_3body_conf}) in the form
\begin{equation}
\label{vetau}
V_{E}^\tau(3:12)=V_{0}^E\tau_{12}Z_{0}(r_{13})Z_{0}(r_{23})\, .
\end{equation}
where the superscript $\tau$ has a meaning that will be soon clarified. 
The radial function $Z_{0}(r)=m_{\pi}^3/(4\pi) z_0(r)$ approaches the Dirac $\delta$-function in the limit of infinite cutoff. Strictly speaking, the local version of $V_E$ is a genuine ``contact term'' in this limit only, while for finite values of the cutoff it  acquires a finite range. 

In addition to $V_{E}^\tau$ of Eq.~(\ref{vetau}), the chiral expansion leads to the appearance of five spin-isopin structures in the contact term. For example, the scalar contribution is
\begin{equation}
V_{E}^I(3:12)=V_{0}^E Z_{0}(r_{13})Z_{0}(r_{23})\, .
\end{equation}
Within this context,  the superscripts $\tau$ and $I$ identify the $\tau_{12}$ and scalar contact terms. 

In Ref. \cite{epelbaum_02} it has been shown that, once the sum over all cyclic permutation is performed, all contributions to the product between the potential and the antisymmetrization  operator $\mathcal{A}_{123}$ have the same spin-isospin structure. Therefore it is convenient to take into account just one of the contact terms. 
This result  was obtained in momentum space, without the cutoff functions $F_\Lambda$. As a consequence,  in coordinate space it only holds true in the limit of infinite cutoff.
In particular,  for $V_{E}^\tau(3:12)$ and $V_{E}^I(3:12)$, it turns out that
\begin{equation}
\sum_{cycl}V_{0}^E\delta(\mathbf{r}_{13})\delta(\mathbf{r}_{23})\tau_{12}\mathcal{A}_{123}
=-\sum_{cycl}V_{0}^E\delta(\mathbf{r}_{13})\delta(\mathbf{r}_{23})\mathcal{A}_{123}\, ,
\label{eq:antisimm}
\end{equation}
making this two terms equivalent. The limit of infinite cutoff is crucial, because the radial part of the exchange operator, when multiplied by the Dirac 
$\delta$-functions, is nothing but the identity
\begin{equation}
e^{i\mathbf{k}_{ij}\cdot \mathbf{r}_{ij}}\delta(\mathbf{r}_{ij})=1\, .
\end{equation}
After the regularization, i.e. with the $\delta$-function replaced by  $Z_0$, the proof is spoiled and the six different structures are no longer equivalent.

In PNM contact terms involving three or more neutrons vanish because of Pauli principle. On the other hand, the expectation value of the contact terms of the NNLOL potential can 
be different from zero.

Let us assume that reproducing the binding energies of light nuclei and $^2a_{nd}$ require a repulsive $V_{E}$ . Then, one has to choose either  $c_{E}^{\tau_{12}}<0$ 
or $c_{E}^{I}>0$. In PNM, as
\begin{equation}
\langle \tau_{12} \rangle_{PNM}=1\,,
\end{equation}
it turns out that $V_{E}^{\tau}$ is attractive and $V_{E}^{I}$ repulsive. This means that fitting the binding energies and the $n-d$ scattering length with either 
$V_{E}^{\tau}$ or $V_E^{I}$ alone leads to an ambiguity in the expectation value of the potential.  

\begin{figure}[!ht]
\begin{center}
\includegraphics[width=6.0cm,angle=270]{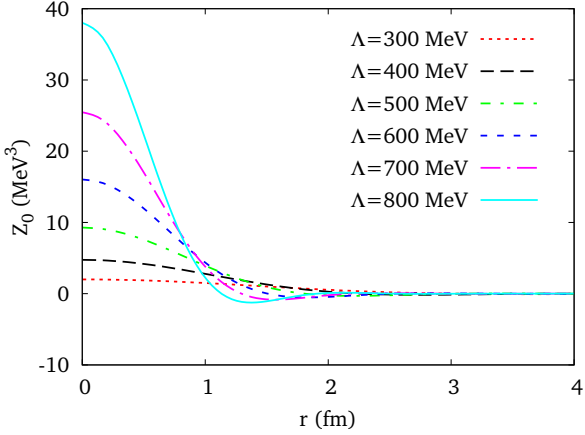}
\caption{Radial dependence of the Function $Z_{0}(r)$, appearing in Eq.(\ref{vetau}), plotted for different values of the cutoff $\Lambda$.}
\end{center}
\end{figure}

By expanding the cutoff function
\begin{equation}
F_\Lambda(q^2)=e^{-q^4/\Lambda^4}\sim 1 -\frac{q^4}{\Lambda^4} +
O\Big(\frac{q^8}{\Lambda^8}\Big)\, ,
\end{equation}
one finds
\begin{align}
V_{E}^{\tau}(3:12)&=V_{0}^E\tau_{12}\Big[\delta(r_{13})\delta(r_{23}
)+O\Big(\frac{q^4}{\Lambda^4}\Big)\Big]\nonumber\\
V_{E}^{I}(3:12)&=V_{0}^E\Big[\delta(r_{13})\delta(r_{23})+O\Big(\frac{q^4}{
\Lambda^4}\Big)\Big]\, ,
\end{align}
implying that in PNM
\begin{equation}
\langle V_{E}^{I,\tau}(3:12)
\rangle_{PNM}=O\Big(\frac{q^4}{\Lambda^4}\Big)\, .
\end{equation}
From the above equation it becomes apparent that the expectation value of the three-nucleon potential, as well as its sign ambiguity, is nothing but a a cutoff effect. Hemce, it should be regarded as  a theoretical uncertainty. Note that, since $\Lambda_\chi\simeq \Lambda$, then $\langle V_{E}\rangle_{PNM}$ is of the same order of the next term in chiral expansion. 

To clarify this issue,  let us consider a simple system: a Fermi gas of neutrons, in which correlations among particles are not present. 
The expectation value of the contact interaction reads
\begin{align}
\frac{\langle V_{E}^{I,\tau} \rangle_{PNM}^{FG}}{A}&=\frac{\rho^2}{2}V_{0}^E\int
d^3r_{12}d^3r_{13}
Z_{0}(r_{12})Z_{0}(r_{13})\times\nonumber\\
&\Big(1-\frac{\ell(r_{12})^2}{2}-\frac{\ell(r_{13})^2}{2}
-\frac{\ell(r_{23})^2}{2}\nonumber\\
&+\frac{\ell(r_{12})\ell(r_{13})\ell(r_{23})}{2}\Big)\, ,
\label{eq:VE_PNM_FG}
\end{align}
where $A$ is the number of neutrons. The factor $1/2$ includes the $1/3!$ arising from the unrestricted sum over particle indices $123$, multiplied by a factor $3$ from the cyclic permutations of the potential, all giving the same contribution. The Slater function $\ell(r_{ij})$, for a system of fermions with degeneracy $d$ is given by
\begin{align}
\ell(r_{ij})&=\frac{d}{N}\sum_{|\mathbf{k}|<k_F}e^{-i
\mathbf{k}_n\cdot\mathbf{r}_i}e^{i \mathbf{k}_n\cdot\mathbf{r}_j} \nonumber\\
&=3\Big[\frac{\sin(k_Fr_{ij})-k_Fr_{ij}\cos(k_Fr_{ij})}{(k_Fr_{ij})^3}\Big]\, .
\end{align}
where $k_F = (\frac{6 \pi^2 \rho}{d})^{1/3}$ is the Fermi momentum. It can be  easily seen that, if $V_{E}^{I,\tau}(1:23)\propto \delta(r_{12})\delta(r_{13})$, then 
\begin{equation}
\frac{\langle V_{E}^{I} \rangle_{PNM}^{FG}}{A}=0\, .
\end{equation}
 
Consider now a Fermi gas with equal numbers of protons and neutrons, where
\begin{align}
\frac{\langle V_{E}^{\tau} \rangle_{SNM}^{FG}}{A}&=\frac{\rho^2}{2}V_{0}^E\int
d^3r_{12}d^3r_{13}
Z_{0}(r_{12})Z_{0}(r_{13})\times\nonumber\\
&\Big(-\frac{3}{4}\ell(r_{23})^2+\frac{3}{8}\ell(r_{12})\ell(r_{
13})\ell(r_{23})\Big)
\label{eq:VEtau_SNM_FG}
\end{align}
and
\begin{align}
\frac{\langle V_{E}^{I} \rangle_{SNM}^{FG}}{A}&=\frac{\rho^2}{2}V_{0}^E\int
d^3r_{12}d^3r_{13}
Z_{0}(r_{12})Z_{0}(r_{13})\times\nonumber\\
&\Big(1-\frac{\ell(r_{12})^2}{4}-\frac{\ell(r_{13})^2}{4}-\frac{\ell(r_{23}
)^2}{4}+\nonumber\\
&\frac{\ell(r_{12})\ell(r_{13})\ell(r_{23})}{8}\Big)\, .
\label{eq:VEI_SNM_FG}
\end{align}
In the limit of infinite cutoff the above equations imply
\begin{align}
\frac{\langle V_{E}^{\tau_{12}} \rangle_{SNM}^{FG}}{A}&=-\frac{3}{16}\rho^2
V_{0}^E\nonumber\\
\frac{\langle V_{E}^{I} \rangle_{SNM}^{FG}}{A}&=\frac{3}{16}\rho^2 V_{0}^E\, .
\label{eq:aympt_SNM}
\end{align}
As expected from Eq.~(\ref{eq:antisimm}), the two contributions have opposite sign.

We have computed the expectation values of Eqs.~(\ref{eq:VE_PNM_FG}), (\ref{eq:VEtau_SNM_FG}) and (\ref{eq:VEI_SNM_FG}) for different values of the cutoff $\Lambda$  
and density $\rho=0.16\,\text{fm}^{-3}$. The results listed  in Table \ref{tab:contact_cut_scan_PNM} show that for PNM the larger the cutoff the smaller the expectation value of the three nucleon contact term. Note that for $\Lambda = 500$ MeV, the expectation value is still sizably different from the asymptotic limit.

As far as SNM is concerned (see Table \ref{tab:contact_cut_scan_SNM}), as the cutoff increases the possible choices of the three nucleon contact term 
tend to the asymptotic values of Eq.~(\ref{eq:aympt_SNM}). As in the case of PNM, the results corresponding to $\Lambda~=~500$~MeV, are significantly
different from the asymptotic values.

\begin{table}[]
\begin{center}
\caption{Cutoff dependence of the expectation values of the three body contact term of the NNLOL potential in noninteracting PNM. \label{tab:contact_cut_scan_PNM}}
\vspace{0.3cm}
\begin{tabular}{c c} 
\hline 
 $\Lambda\,\text{(MeV)}$ & $
\langle V_{E}^{I,\tau_{12}} \rangle_{PNM}^{FG}/A \,\text{(MeV)}$\\ 
\hline
$300$ &  9.15 \\ 
$400$ &  5.95  \\ 
$500$ &  3.60  \\ 
$600$ &  2.15  \\  
$700$ &  1.30  \\ 
$800$ &  0.81  \\ 
\hline
$\infty$ & 0\\ 
\hline
\end{tabular} 
\vspace{0.1cm}
\end{center}
\end{table}

\begin{table}[]
\begin{center}
\caption{Same as in Table \ref{tab:contact_cut_scan_PNM}, but for SNM.\label{tab:contact_cut_scan_SNM}}
\vspace{0.3cm}
\begin{tabular}{c c c } 
\hline 
 $\Lambda\,\text{(MeV)}$ & $ \langle V_{E}^{\tau_{12}} \rangle_{SNM}^{FG}/A
\,\text{(MeV)}$ & $ \langle V_{E}^{I} \rangle_{SNM}^{FG}/A \,\text{(MeV)}$\\ 
\hline
$300$ & -2.61 & 10.21  \\ 
$400$ & -3.61 & 8.15   \\ 
$500$ & -4.37 & 6.93   \\ 
$600$ & -4.87 & 6.30   \\  
$700$ & -5.15 & 5.98   \\ 
$800$ & -5.30 & 5.81   \\ 
\hline
$\infty$ & -5.55 & 5.55 \\ 
\hline
\end{tabular} 
\vspace{0.1cm}
\end{center}
\end{table}

We emphasize that the parameter $c_E$ has not been included in this analysis, even though it is  itself cutoff dependent. Unfortunately, the authors of 
Ref.~\cite{kievsky_10} kept $\Lambda$ fixed to $500\,\text{MeV}$. Had this not been the case, their fit to the experimental data would have 
resulted in a set of different constants $c_E$, corresponding to different values of $\Lambda$. It would have been interesting to extrapolate 
the expectation value of $V_E$ to the limit of infinite $\Lambda$, where the cutoff effects associated with the regularization procedure are expected to vanish. 



\subsection{FHNC/SOC calculations}
\begin{figure}[!ht]
\includegraphics[width=8.0cm]{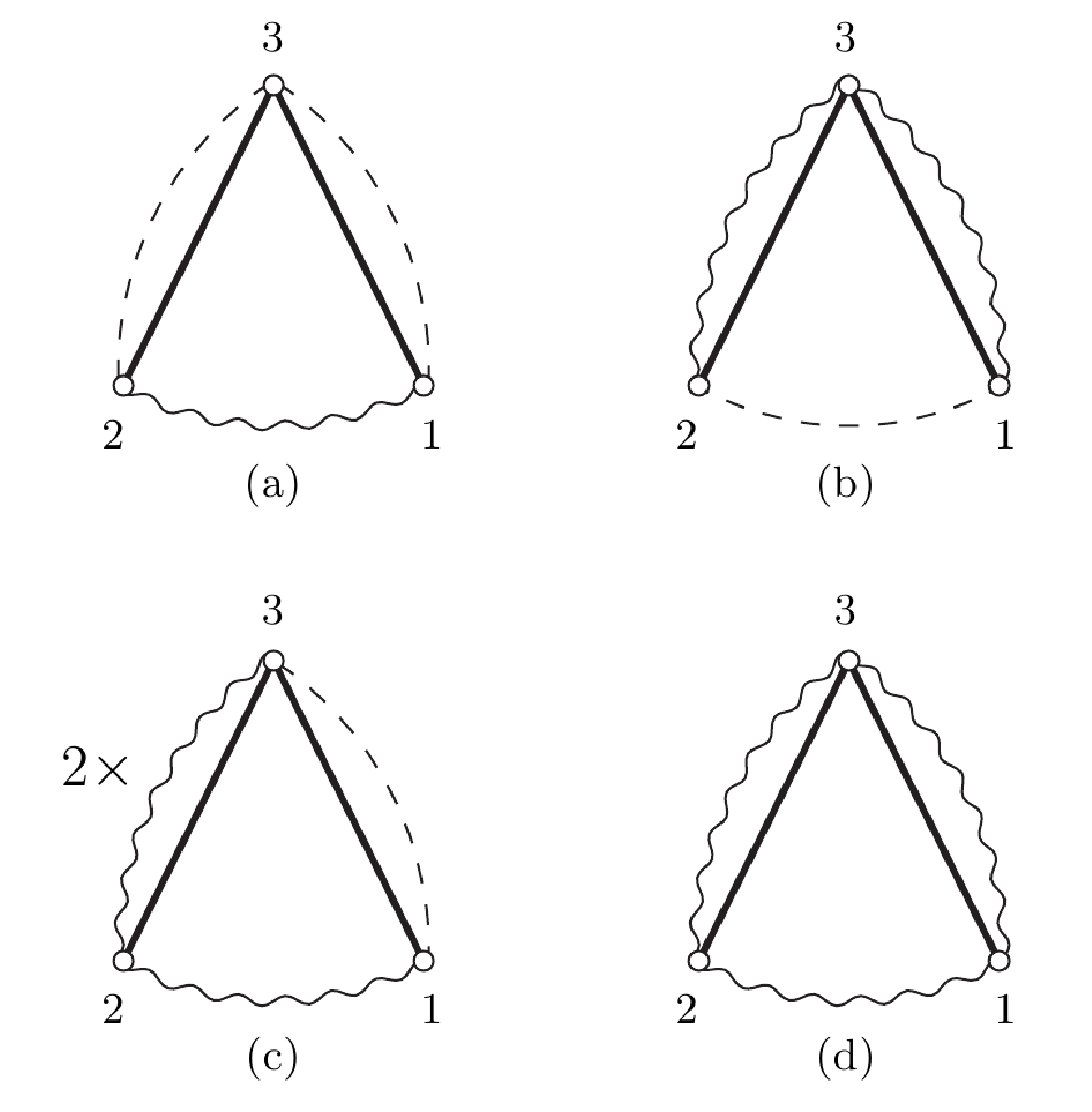}
\vspace{0.3cm}
\caption{Cluster diagrams contributing to the expectation value of  $V^{2\pi}$.}
\label{fig:tbp_2pi}
\end{figure}

\begin{figure}[!ht]
\includegraphics[width=8.0cm]{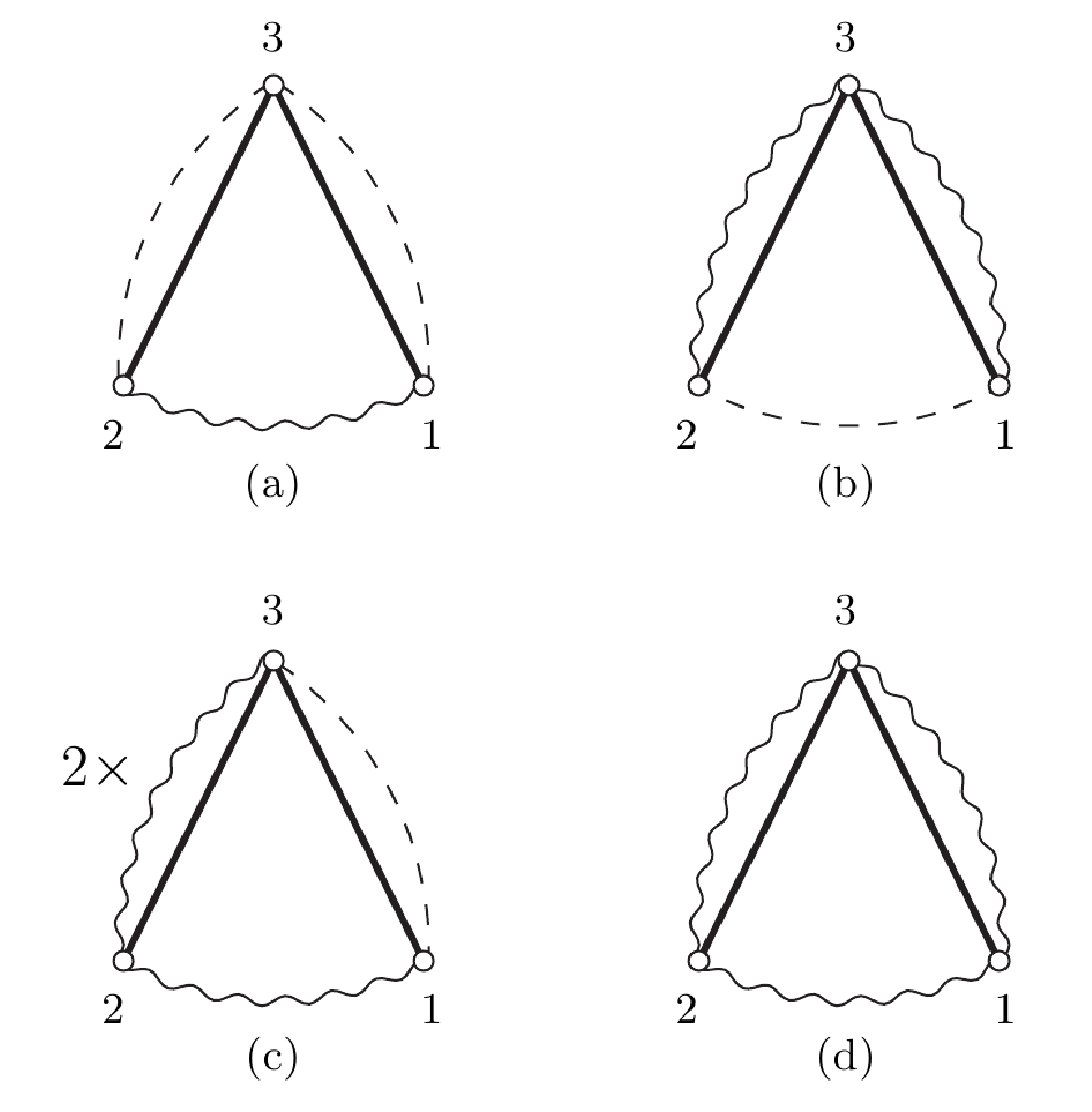}
\vspace{0.3cm}
\caption{Same as in Fig. \ref{fig:tbp_2pi}, but for  $V^{R}$.\label{fig:tbp_scalar}}
\end{figure}

The diagrams involved in the FHNC/SOC calculation of the expectation values of the $V^{2\pi}$ and $V^R$ terms of the UIX potential are depicted in Figs. \ref{fig:tbp_2pi} and  \ref{fig:tbp_scalar}, respectively. The thick lines represent the potential, while dashed and wavy lines correspond to generalized scalar and operatorial correlations, denoted by $Z^c$ and $Z^p$ in Ref.~\cite{carlson_83}. Double wavy lines represent Single Operator Rings (SOR), while vertex corrections, although included in the calculations, are not shown. The definitions of all these quantities can be found in Refs.\cite{wiringa_78,pandharipande_79}. 

Note that, because of the symmetry properties of the wave function, we can restrict our analysis to the permutation $(3:12)$. Taking into account the other permutations 
results in the appearance of a multiplicative factor.
 
The computation of all of diagrams (3.a), (3.b) and (3.c) and all diagrams of Fig. \ref{fig:tbp_scalar} is outlined in Ref.~\cite{carlson_83},  while the contribution 
of digram (3.d), involving three non central correlations was first taken into account by the authors of Ref.~\cite{wiringa_88}.

Using the relations for the constants and the radial functions given in Eqs. (\ref{eq:const_rel}) and (\ref{eq:rad_gen_UIX}), the computation of 
the diagrams of Fig. \ref{fig:tbp_2pi} with the $V_3$ and $V_4$ terms of both the TM$^\prime$ and NNLOL potentials is the same as that of 
$\hat{V}^{2\pi}$ reported in Ref.~\cite{carlson_83}.

Thanks to the identity
 \begin{align}
&(\boldsymbol{\sigma}_1\cdot\hat{r}_{13})(\boldsymbol{\sigma}_2\cdot\hat{r}_{23})(\hat{r}_{13}\cdot\hat{r}_{23})=\nonumber\\
&\qquad \frac{1}{18}\{\sigma_{13}+S_{13},\sigma_{23}+S_{23}\}\, ,
\end{align}
the term $V_1$ of Eq.~(\ref{eq:2pi_3body_conf}), appearing in both the TM$^\prime$ and the NNLOL potentials, can be written in the form
\begin{align}
V_{1}(3:12)&=\frac{W_0}{36}\{\tau_{13},\tau_{23}\}\{\sigma_{13}+S_{13},\sigma_{23}+S_{23}\}\times\nonumber\\
&\frac{r_{13}r_{23}}{(\hat{r}_{13}\cdot \hat{r}_{23})}y(r_{13})y(r_{23})\, .
\end{align}
Aside from the radial function, $V_1$ is completely equivalent to $V_3$, the anticommutator term of the UIX potential. Therefore, we were 
allowed to use again the results of  Ref.~\cite{carlson_83}.

Furthermore, exploiting the identities
\begin{align}
(\boldsymbol{\sigma}_1\cdot\vec{r}_{23})(\boldsymbol{\sigma}_2\cdot\vec{r}_{23})
&=\frac{r_{23}^2}{6}\{S_{23}+\sigma_{23},\sigma_{13}\}\nonumber\\
(\boldsymbol{\sigma}_2\cdot\vec{r}_{13})(\boldsymbol{\sigma}_1\cdot\vec{r}_{13})
&=\frac{r_{13}^2}{6}\{\sigma_{23},S_{13}+\sigma_{13}\}\, ,
\end{align}
we can rewrite the $V_D$ term in a form that has again the same spin-isospin structure as the anticommutator contribution of the UIX potential
\begin{align}
V_{D}(3:12)&=\frac{W_{0}^D}{4}\{\tau_{13},\tau_{23}\}[\{\sigma_{13},\sigma_{23}\}V_{D}^{YY}(r_{13},r_{23})+\nonumber\\
&\qquad \{S_{13},\sigma_{23}\}V_{D}^{TY}(r_{13},r_{23})+\nonumber\\
&\qquad\{\sigma_{13},S_{23}\}V_{D}^{YT}(r_{13},r_{23})]\, ,
\label{eq:V_D_rearrange}
\end{align}
where
\begin{align}
V_{D}^{YY}({r}_{13},{r}_{23})&=Y(r_{13})z_{0}(r_{23})+z_{0}(r_{13})Y(r_{23})\nonumber\\
V_{D}^{YT}({r}_{13},{r}_{23})&=z_0(r_{13})T(r_{23})\nonumber\\
V_{D}^{TY}({r}_{13},{r}_{23})&=T(r_{13})z_0(r_{23})\, .
\end{align}
In conclusion, including $V_D$ amounts to properly adding the above radial functions to those already appearing in $V_3$. 

The $V_E$ term of TM$^\prime$ is completely equivalent to $V_R$ (see Eq.~(\ref{eq:E_equiv}) ). This allowed us to use the results of  Ref.~\cite{carlson_83} for 
the diagrams of Fig. \ref{fig:tbp_scalar}. The same holds true for the chiral contact term $V_E$ in PNM, as $\langle\tau_{ij}\rangle_{PNM}=1$,  
while in SNM the calculation of $V_E$ requires the evaluation of the diagrams of Fig. \ref{fig:tbp_2pi}.

The expression of diagram (\ref{fig:tbp_2pi}.a) is
\begin{align}
(\ref{fig:tbp_2pi}.a)=&\frac{c_E}{2}\rho^2\sum_{ex}\sum_{p}\int d^3r_{12}d^3r_{13}Z_{xy,13}^{c}Z_{x'y',23}^{c}\times\nonumber\\
&\qquad Z_{x''y'',12}^{p}\text{C}\Big(V_{E}(3:12)O_{12}^{p}\Big) .
\label{eq:dia21_2}
\end{align}
As pointed out  in Ref.~\cite{carlson_83}, integrating and tracing over the radial and spin-isospin variables of particle $3$ leads to the appearance of an effective density dependent interaction 
\begin{align}
\sum_p V_{E,yy',12}^p(\rho) O_{12}^{p}&=\rho\sum_{ex}\int d^3r_{13}Z_{xy,13}^{c}Z_{x'y',23}^{c}\times\nonumber\\
&\qquad \text{C}_{3}\Big(V_{E}(3:12)\Big)\, ,
\label{eq:dia21_3}
\end{align}
such that
\begin{align}
(\ref{fig:tbp_2pi}a)=
&\frac{c_E}{2}\rho\sum_{ex}\sum_{p}\int d^3r_{12}A^pZ_{x''y'',12}^{p}V_{E,yy',12}^p(\rho)\, .
\label{eq:dia21_4}
\end{align}
The subscripts $xy$ label exchange patterns at the ends of the generalized correlation lines. In particular, $dd$ correspond stands for 
direct-direct, $de$ for direct-exchange, $ee$ for exchange-exchange, and $cc$ for incomplete circular exchange. 
The matrix $A_p$ is defined through \cite{pandharipande_79} 
\begin{equation}
C(O^{p}_{ij}O^{q}_{ij})=\delta^{pq}A^p\, ,
\end{equation}
implying $A^p=1,3,3,9,6,18$ for $p=1,6$.

It turns out that the only nonvanishing term of the density dependent potential is
\begin{align}
V_{E\,yy',12}^{\tau}(\rho)&=W_{0}^E\rho\sum_{ex}\int d^3r_{13}Z_{xy,13}^{c}Z_{x'y',23}^{c}\times\nonumber\\
&\qquad z_0(r_{13})z_0(r_{23})\, .
\end{align}

The contribution of diagram (\ref{fig:tbp_2pi}.b) is given by 
\begin{align}
(\ref{fig:tbp_2pi}.b)=&\frac{c_E}{2}\rho^2\sum_{ex}\sum_{p,p'}\int d^3r_{12}d^3r_{13}Z_{xy,13}^{p}Z_{x'y',23}^{p'}\times\nonumber\\
&\qquad Z_{x''y'',12}^{c}\text{C}\Big(V_{E}(3:12)\frac{1}{2}\{O_{13}^{p},O_{23}^{p'}\}\Big)\, .
\label{eq:dia22_1}
\end{align}
From the above expression, it clearly follows that only $\tau$-type generalized correlation lines contribute. Hence
\begin{align}
(\ref{fig:tbp_2pi}b)=&\frac{3}{2} c_EW_{0}^E\rho^2\sum_{ex}\int d^3r_{12}d^3r_{13}Z_{xy,13}^{\tau}Z_{x'y',23}^{\tau}\nonumber\\
&\qquad Z_{x''y'',12}^{c}\,z_0(r_{13})z_0(r_{23})\, .
\end{align}

Diagram (\ref{fig:tbp_2pi}.c) does not contribute to $V_E$, while for diagram (\ref{fig:tbp_2pi}.d), with three  generalized operatorial correlation, 
we find 
\begin{align}
(\ref{fig:tbp_2pi}d)=&\frac{c_E}{2}\rho^2\sum_{ex}\sum_{p,p',p''}\int d^3r_{12}d^3r_{13}Z_{xy,12}^{p}Z_{x'y',13}^{p'}\times\nonumber\\
&\qquad Z_{x''y'',23}^{p''}\text{C}\Big[\frac{V_{E}(3:12)}{3!}(O_{12}^{p}\{O_{13}^{p'},O_{23}^{p''}\}+\nonumber \\
&\qquad O_{13}^{p'}\{O_{12}^{p},O_{23}^{p''}\}+ O_{23}^{p''}\{O_{12}^{p},O_{13}^{p'}\})\Big]\, .
\label{eq:diag24_1}
\end{align}
The calculation of the spin-isospin traces yields
\begin{align}
(\ref{fig:tbp_2pi}d)=&\frac{c_E}{2} W_{0}^E\rho^2\int d^3r_{12}d^3r_{13}\Big(-2Z_{12}^{\tau} Z_{13}^{\tau}Z_{23}^{\tau}+\nonumber\\
&9Z_{12}^{\sigma\tau}Z_{13}^{\sigma}Z_{23}^{\sigma}+9Z_{12}^{\sigma}Z_{13}^{\sigma\tau}Z_{23}^{\sigma\tau}-\nonumber\\
&6Z_{12}^{\sigma\tau}Z_{13}^{\sigma\tau}Z_{23}^{\sigma\tau}+18\xi^{\sigma tt}_{231}Z_{12}^{t}Z_{13}^{t\tau}Z_{23}^{\sigma\tau}-\nonumber\\
&12\xi^{\sigma tt}_{231}Z_{12}^{t\tau}Z_{13}^{t\tau}Z_{23}^{\sigma\tau}+18\xi^{t\sigma t}_{231} Z_{12}^{t}Z_{13}^{\sigma\tau}Z_{23}^{t\tau}+\nonumber\\
&18\xi^{t\sigma t}_{231}Z_{12}^{t\tau}Z_{13}^{\sigma}Z_{23}^{t}-12\xi^{t\sigma t}_{231}Z_{12}^{t\tau}Z_{13}^{\sigma\tau}Z_{23}^{t\tau}+\nonumber\\
&9\xi^{tt\sigma}_{231}Z_{12}^{\sigma}Z_{13}^{t\tau }Z_{23}^{t\tau}+9\xi^{tt\sigma}_{231} Z_{12}^{\sigma\tau}Z_{13}^{t}Z_{23}^{t}-\nonumber\\
&6\xi^{tt\sigma}_{231}Z_{12}^{\sigma\tau}Z_{13}^{t\tau}Z_{23}^{t\tau}-12\xi^{ttt}_{231}Z_{12}^{t\tau}Z_{13}^{t\tau}Z_{23}^{t\tau}+\nonumber\\
&18\xi^{ttt}_{231}Z_{12}^{t}Z_{13}^{t\tau}Z_{23}^{t\tau}+18\xi^{ttt}_{231}Z_{12}^{t\tau}Z_{13}^{t}Z_{23}^{t}+\nonumber \\
&18\xi^{\sigma tt}_{231} Z_{12}^{t\tau}Z_{13}^{t}Z_{23}^{\sigma}\Big)z_0(r_{13})z_0(r_{23})\, .
\end{align}
The matrices $\xi_{231}^{pqr}$, depending on the angles formed by the vectors $\mathbf{r}_1$, $\mathbf{r}_2$ and $\mathbf{r}_3$, are defined in Ref.~\cite{pandharipande_79}.
Following Ref.~ \cite{wiringa_88} we have considered only the direct term of the generalized operatorial correlations. As a consequence, in the previous equation $Z_{ij}^p=Z_{dd,ij}^p$.

In order to find the optimal values of the variational parameters, we have employed a procedure similar to simulated annealing, the details of which are explained in 
Ref.~\cite{lovato_11}. 

The authors of Ref.~\cite{lovato_11} constrained the difference between the Pandharipande-Bethe (PB) and the Jackson-Feenberg (JF) kinetic energies to 
be less than $10\, \%$ of the Fermi Energy $T_F$ and required the sum rule involving the scalar two-body distribution function, $g^c(r_{12})$, to be 
fulfilled with a precision of $3\, \%$. In variational calculations of SNM we have imposed the further condition, firstly considered in Ref.~\cite{wiringa_88}, that 
the sum rule of the isospin component of the two-body distribution function
\begin{equation}
\rho\int d\vec{r}_{12} g^{\tau}(r_{12}) = -3\, ,
\end{equation} 
be also satisfied to the same accuracy.

Using also the sum rules for the spin and spin-isospin two body distribution functions  leads to a sizable increase of the variational energies, which turn out to 
 be much higher than those obtained releasing the additional constraints, as well as the AFDMC results. The same pattern is observed in the results of variational 
 calculations not including TNF.
For this reason, we have enforced the fulfillment of the sum rules for $g^c(r_{12})$ and $g^{\tau}(r_{12})$ only. 

For potentials other than UIX, it turns out that the variational energies of PNM resulting from our optimization procedure are lower than the AFDMC values at  $\rho > \rho_0$ . 
By carefully analyzing the contributions of the cluster expansion diagrams, we realized that the value of diagram (\ref{fig:tbp_2pi}b) was unnaturally large. 
In particular, we have found that a small change in the variational parameters leads to a huge variation of the value of the diagram. Moreover, the minimum of 
the energy in parameter space was reached in a region where the kinetic energy difference was very close to the allowed limit.

To cure this pathology, we have constrained the PB-JF kinetic energy difference to be less than $1$ MeV, regardless of density. The variational energies obtained imposing this new constraint are always larger than the corresponding AFDMC values and the value of diagram (\ref{fig:tbp_2pi}.b) is brought under control. For the sake of consistency, the same constraint on the kinetic energies has been also applied to SNM.

\subsection{AFDMC calculations}
We have computed the EoS of PNM using the AFDMC  approach \cite{schmidt_99} with the TM$^\prime$ and NNLOL chiral potentials combined with the Argonne $v_{8}'$
NN interaction.

An efficient procedure to perform AFDMC calculations with three-body potentials is described in Ref.~\cite{pederiva_04}. Since $V_3$ is equivalent to the anticommutator term 
of the UIX model (while the commutator, $V_4$, is zero in PNM), and in PNM the $V_E$ terms of both the TM$^\prime$ and NNLOL potentials do not show any formal difference 
with respect to the repulsive term of UIX, the inclusion of these terms reduces to a replacement of constants and radial functions. The authors of Ref.~\cite{pederiva_04} 
also described how to handle the $V_1$ for the TM model, and no further difficulties arise in the case of the NNLOL potential.

As  the $V_D$ term has never been encompassed in AFDMC, it is worthwhile showing how the calculation of this term reduces to a matrix multiplication. 
The expectation value of $V_D$ is given by
\begin{equation}
\langle V_D \rangle=\sum_{i<j<k}[V_D(i:jk)+V_D(j:ik)+V_D(k:ij)]
\end{equation}
with $V_D(i:jk)=V_D(i:kj)$ (otherwise all six permutations need to be summed). Thanks to this property one can write
\begin{equation}
\langle V_D \rangle=\sum_{i<k,j}V_D(j:ik)\, .
\end{equation}
It is possible to write $V_D(j:ik)$ of Eq.~(\ref{eq:V_D_rearrange}) in terms of cartesian components operators
\begin{align}
V_D(j:ik)=&(Y_{\alpha i;\beta j}Z_{\gamma j;\delta k}+Z_{\alpha i;\beta j}Y_{\gamma j;\delta k}+\nonumber \\
&\,T_{\alpha i;\beta j}Z_{\gamma j;\delta k}+Z_{\alpha i;\beta j}T_{\gamma j;\delta k})\times\nonumber\\
&\,\{\sigma_{i}^\alpha\sigma_{j}^\beta,\sigma_{j}^\gamma\sigma_{k}^\delta\}\, ,
\end{align}
where
\begin{align}
Y_{\alpha i;\beta j}&=Y(r_{ij})\delta^{\alpha\beta}\nonumber\\
Z_{\alpha i;\beta j}&=z_0(r_{ij})\delta^{\alpha\beta}\nonumber \\
T_{\alpha i;\beta j}&=T(r_{ij})(3\hat{r}_{ij}^\alpha \hat{r}_{ij}^\beta-\delta^{\alpha\beta})\, .
\end{align}
The anticommutation relation $\{\sigma_{i}^\alpha,\sigma_{j}^\beta\}=2\delta^{\alpha\beta}$ makes the expectation value of $V_D$ a sum of $3N\times3N$ matrix multiplications
\begin{align}
\langle V_D \rangle&=2\sum_{i<k,j}(Y_{\alpha i;\beta j}Z_{\beta j;\delta k}+Z_{\alpha i;\beta j}Y_{\beta j;\delta k}+\nonumber \\
&\,T_{\alpha i;\beta j}Z_{\beta j;\delta k}+Z_{\alpha i;\beta j}T_{\beta j;\delta k})\sigma_{i}^\alpha\sigma_{k}^\delta\,\nonumber\\
&=2\sum_{i<k}(\{Y,Z\}+\{T,Z\})_{\alpha i;\delta k}\,\sigma_{i}^\alpha\sigma_{k}^\delta\, ,
\end{align}
analogous to those of Ref.~\cite{pederiva_04}. In order to compute the expectation value of $V_D$ the former expression has been added to the cartesian matrices 
associated with the two-body potential.

Following Ref.~\cite{lovato_11}, we simulated PNM with $A=66$ neutrons in a periodic box, as described in Refs.~\cite{gandolfi_08,gandolfi_09}, using the fixed-phase approximation. For $66$ neutrons finite-size effects on the kinetic energy have been found to be small, as its value is very close to the thermodynamic limit. Moreover, as shown in Ref.~\cite{gandolfi_09}, the energy per particle obtained with $66$ neutrons imposing the Periodic Box Condition (PBC) differs by no more than 2\% from the asymptotic value calculated with Twist Averaged Boundary Conditions (TABC). 

Finite-size effects are expected to be larger when the density is bigger, as the dimension of the box decreases. In order to check the validity 
of our calculations, at $\rho=0.48\,\text{fm}^{-3}$  we have repeated the calculation with $114$ neutrons. 
For all potentials, no significant deviations from the energy per particle obtained with $66$ neutrons have been observed.

\section{Nuclear Matter EoS }
\label{results}
\subsection{TM$^\prime$ potential}

\begin{figure}[!ht]
\centering
\includegraphics[width=6.2cm,angle=270]{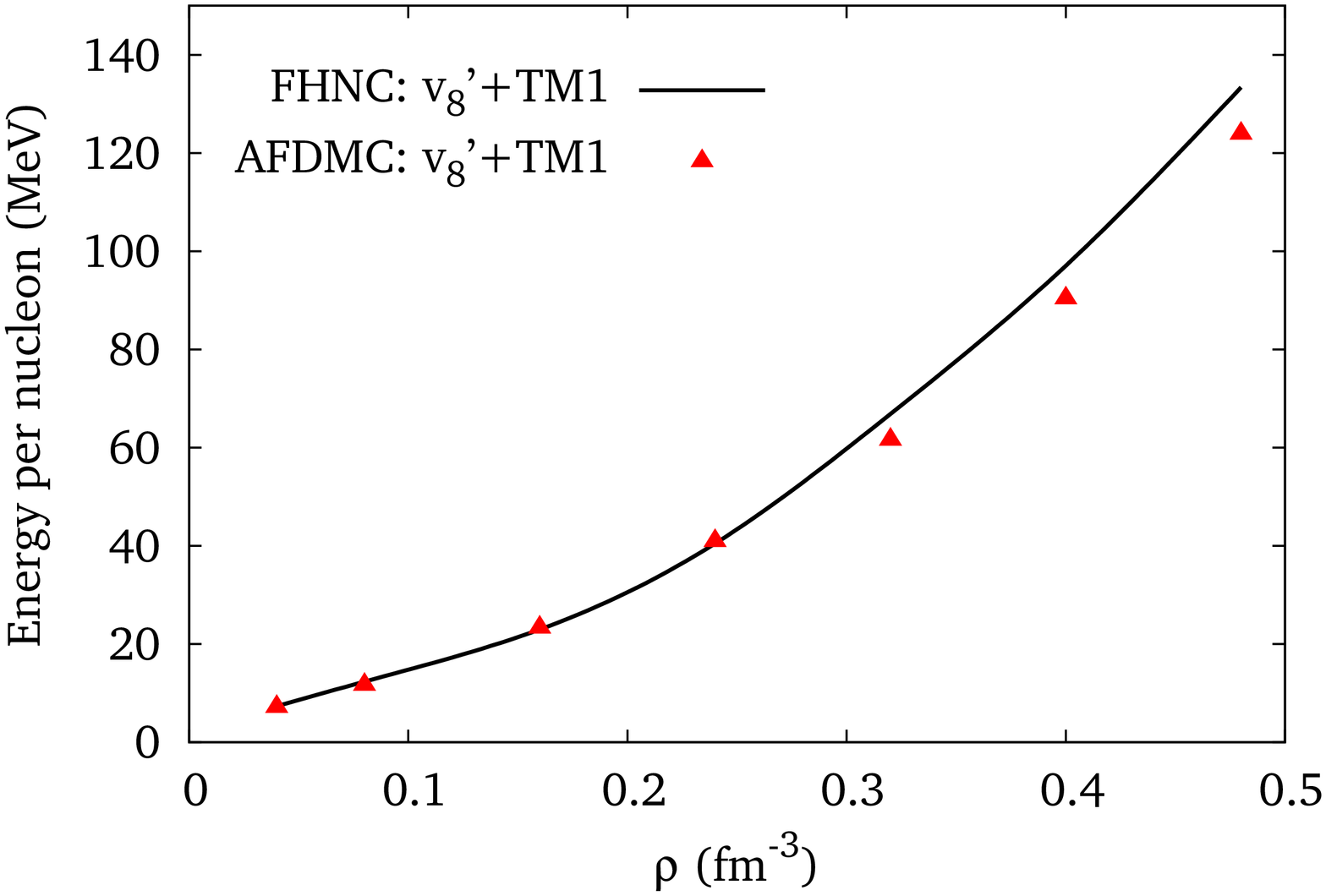}
\hspace{5mm}
\includegraphics[width=6.2cm,angle=270]{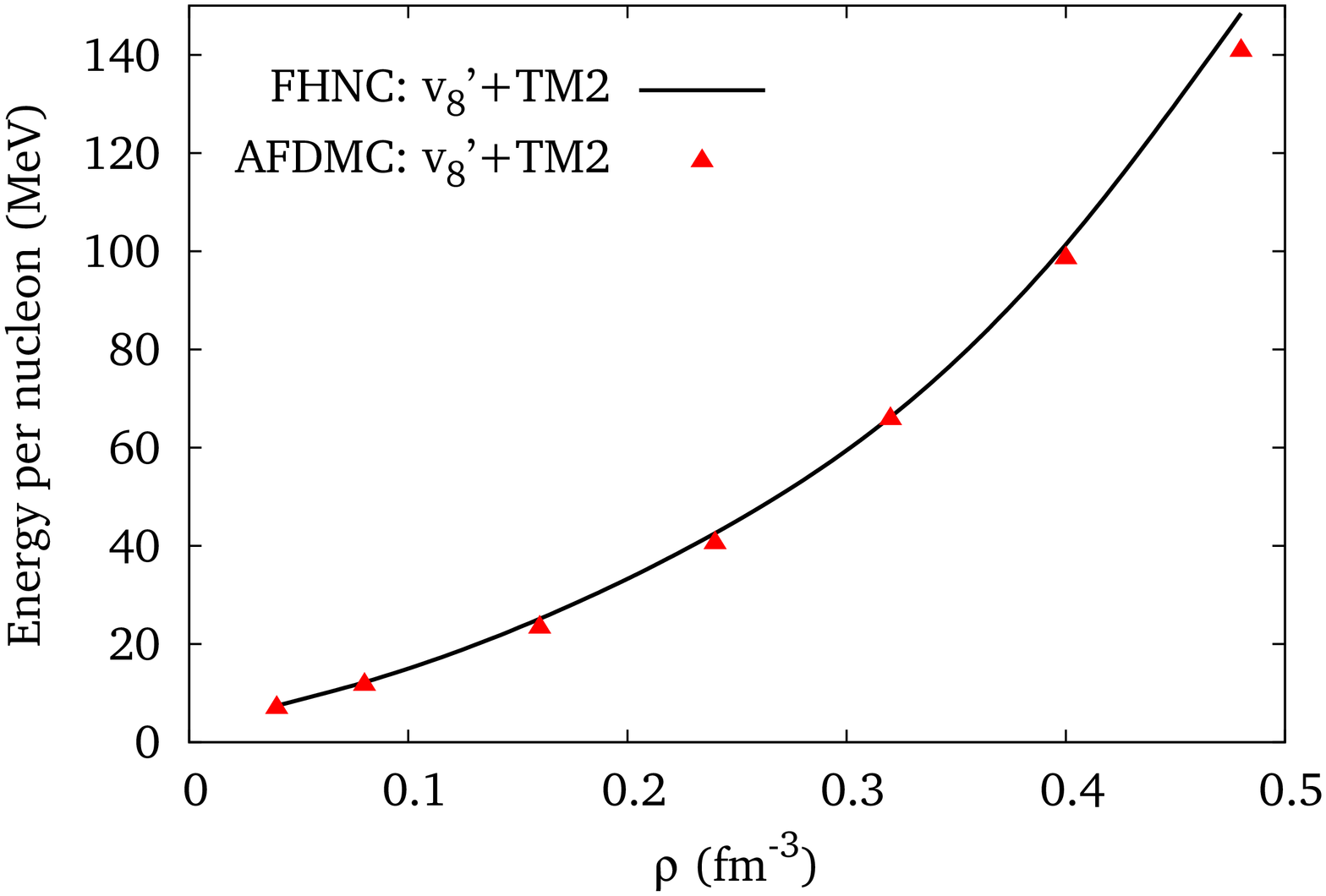}
\hspace{5mm}
\includegraphics[width=6.2cm,angle=270]{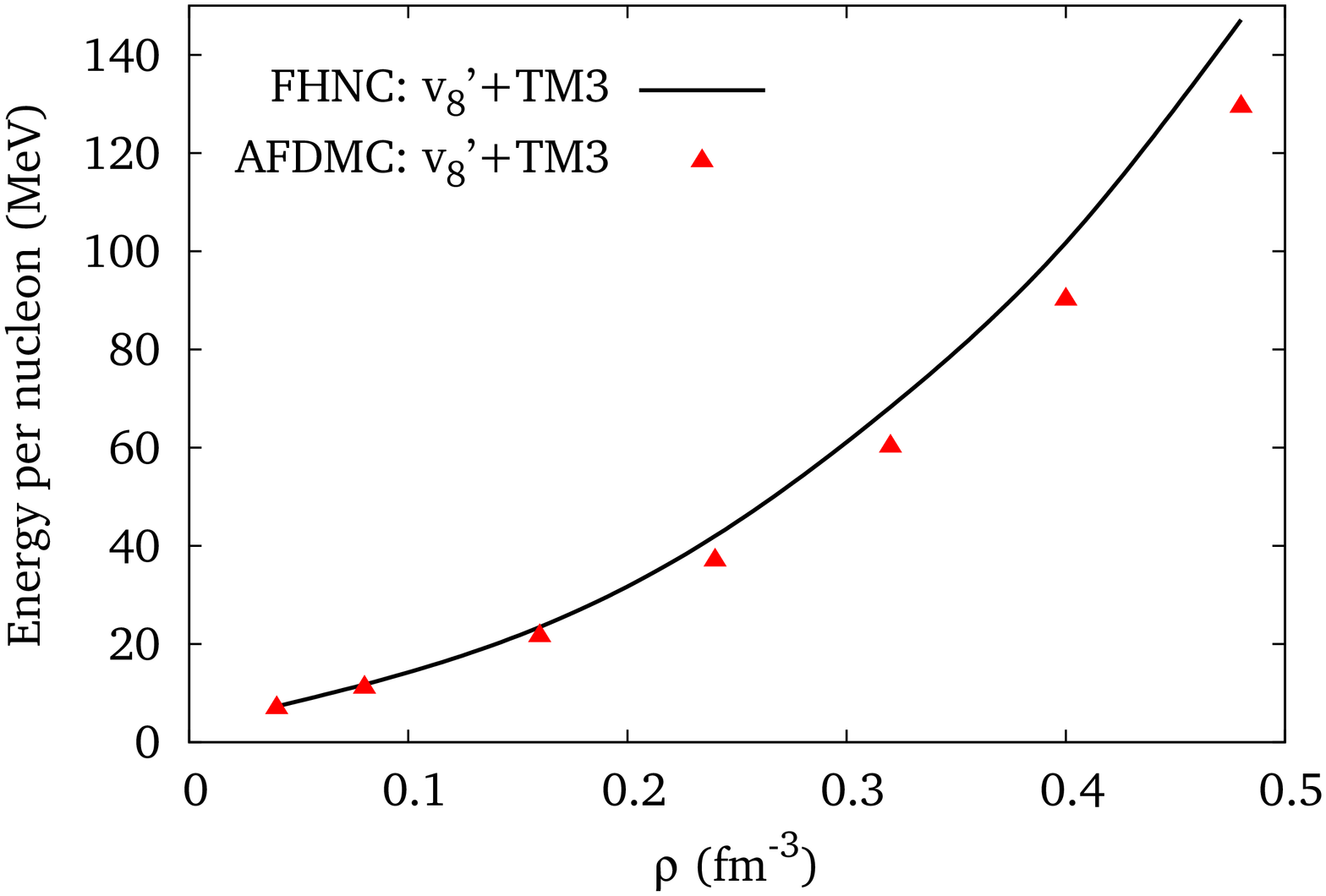}
\caption{Equation of state of PNM obtained using the AFDMC (triangles) and FHNC/SOC (solid lines) approaches with the TM$^\prime$ plus $v_8^\prime$ hamiltonian.\label{fig:tm_compare_PNM}}
\end{figure}

The results of Fig. \ref{fig:tm_compare_PNM}, showing the density dependence of the energy per nucleon in PNM, indicate that,  
once the new constraint on the difference between PB and JF kinetic energies is imposed, the agreement between FHNC/SOC (solid line) and AFDMC (triangles) results 
is very good. 

\begin{figure}[!h]
\vspace{0.2cm}
\begin{center}
\includegraphics[angle=270,width=9.0cm]{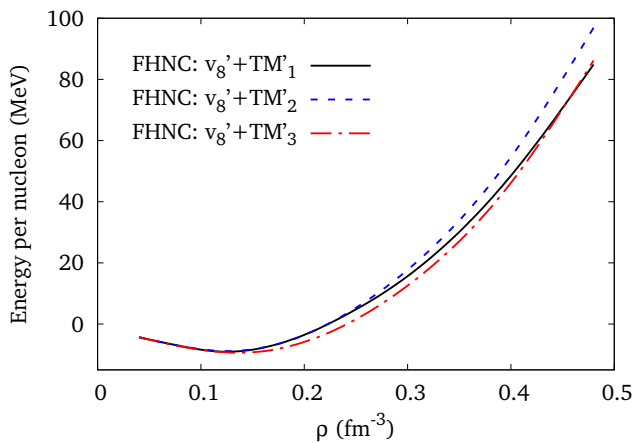}
\caption{Equation of state of SNM resulting from FHNC/SOC variational calculations with the TM$^\prime$ plus $v_8^\prime$ hamiltonian.\label{fig:tm_snm}}
\end{center}
\end{figure}

The most striking feature of the results displayed in Fig. \ref{fig:tm_snm} is that, despite the parameters of the three body potentials being different, all SNM EoS obtained from the TM$^\prime$ potential turn out to be very close to each other. This is probably due to the fact that these potentials are designed to reproduce not only the binding energies of $^3$H and $^4$He, but also the n-d doublet scattering length $^2a_{nd}$.

It is remarkable that although the parameters of TM$^\prime$ potentials were not adjusted to reproduce nuclear matter properties, the EoS saturates at densities only slightly lower than 
$\rho_0=0.16\text{fm}^{-3}$, and the compressibilities are in agreement with the experimental value $K\approx 240\,\text{MeV}$. 
On the other hand, the binding energies are larger than the empirical value $E_0=-16\,\text{MeV}$ and rather close to the one obtained from the UIX potential,  $\sim 10\,\text{MeV}$ \cite{lovato_11}. The numerical values of all these quantities are listed in Table \ref{table:parameters_tm}.

\begin{table}[!h]
\caption{Saturation density, binding energy per particle and compressibility of SNM corresponding to the TM$^\prime$ EoS displayed in Fig. \ref{fig:tm_snm}. \label{table:parameters_tm}}
\vspace{0.3cm}
\begin{tabular}{ c c c c} 
\hline 
 	& TM$^\prime_1$ & TM$^\prime_2$ & TM$^\prime_3$\\
\hline
$\rho_0$ (fm$^{-3}$)&  0.12  &  0.13  &  0.14  \\

$E_0$ (MeV) &  -9.0  &  -8.8  &  -9.4 \\

K (MeV) &  266  &  243  &  249   \\ 
\hline
 
\end{tabular} 
\vspace{0.1cm}
\end{table}
\subsection{NNLOL chiral potentials}
\label{sottosezione}

\begin{figure*}[!t]
\centering
\includegraphics[width=6.2cm,angle=270]{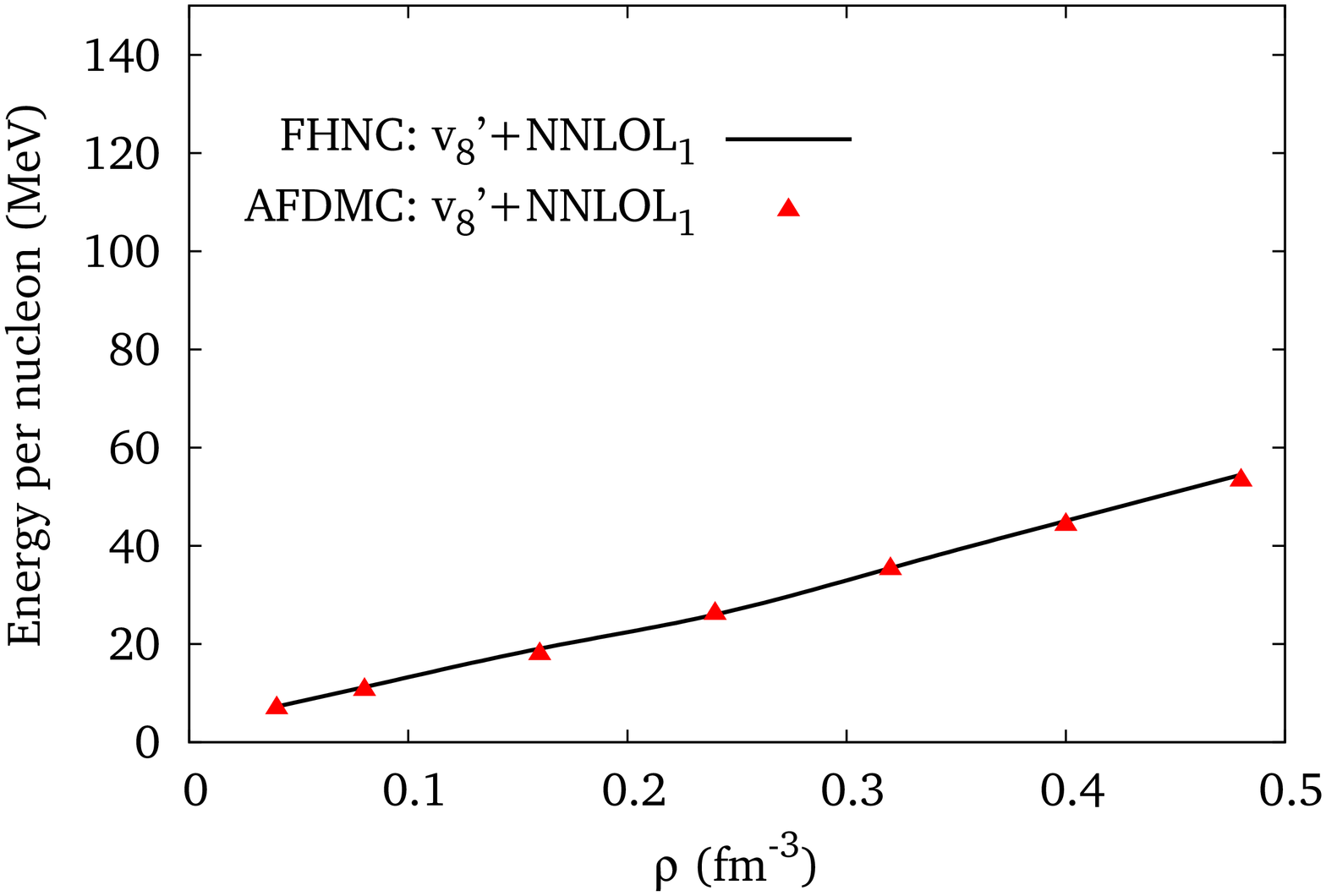}
\includegraphics[width=6.22cm,angle=270]{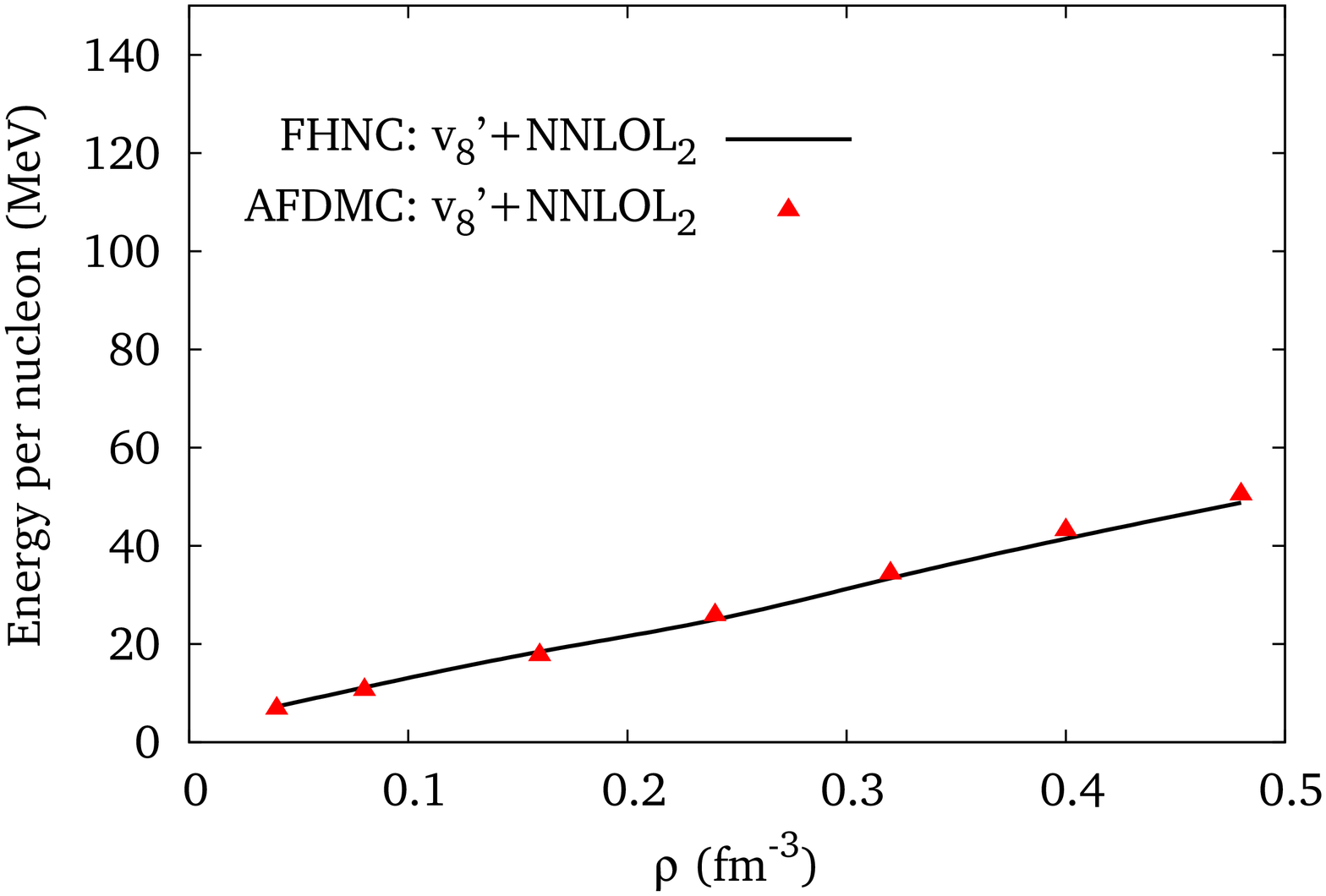}
\includegraphics[width=6.2cm,angle=270]{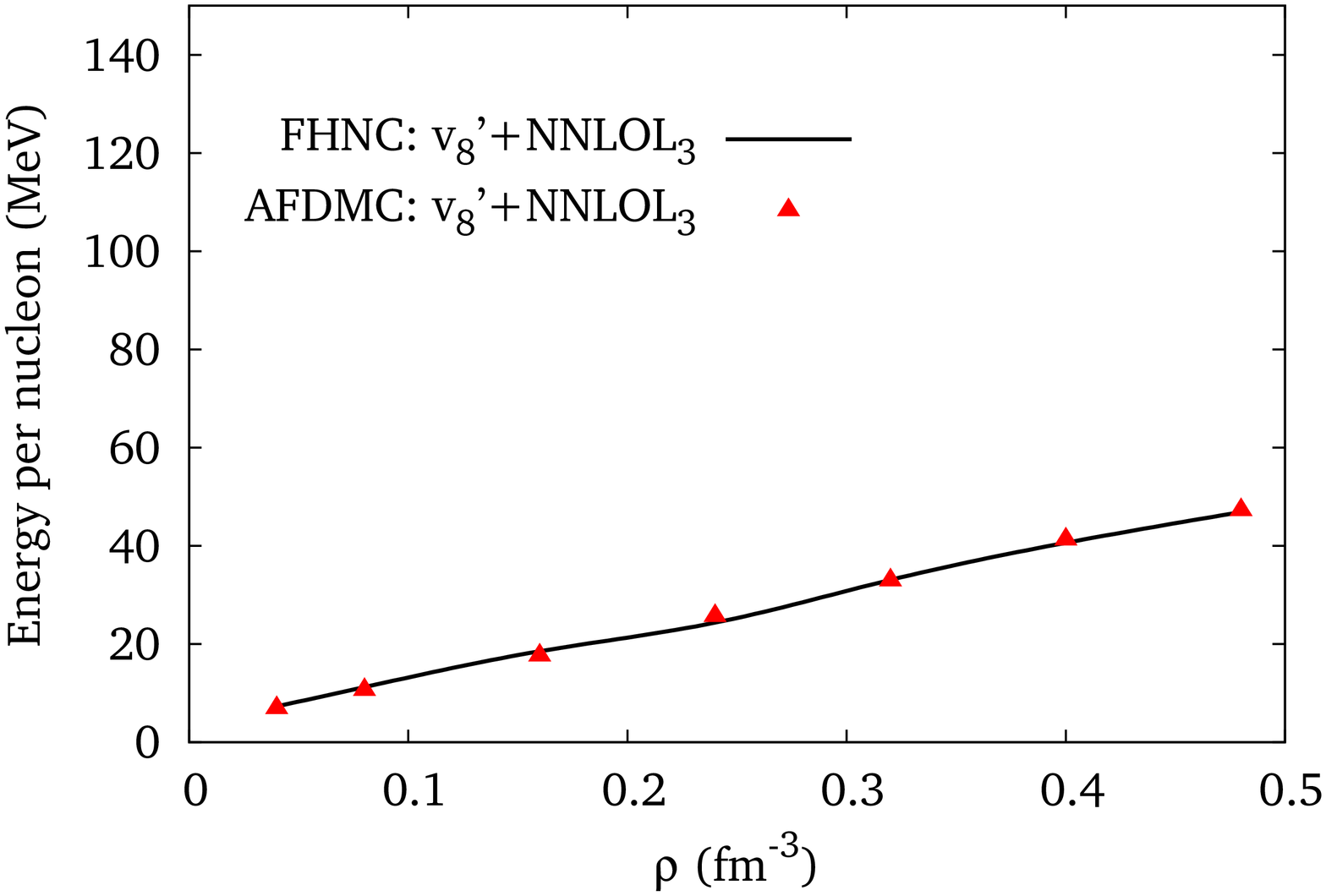}
\includegraphics[width=6.2cm,angle=270]{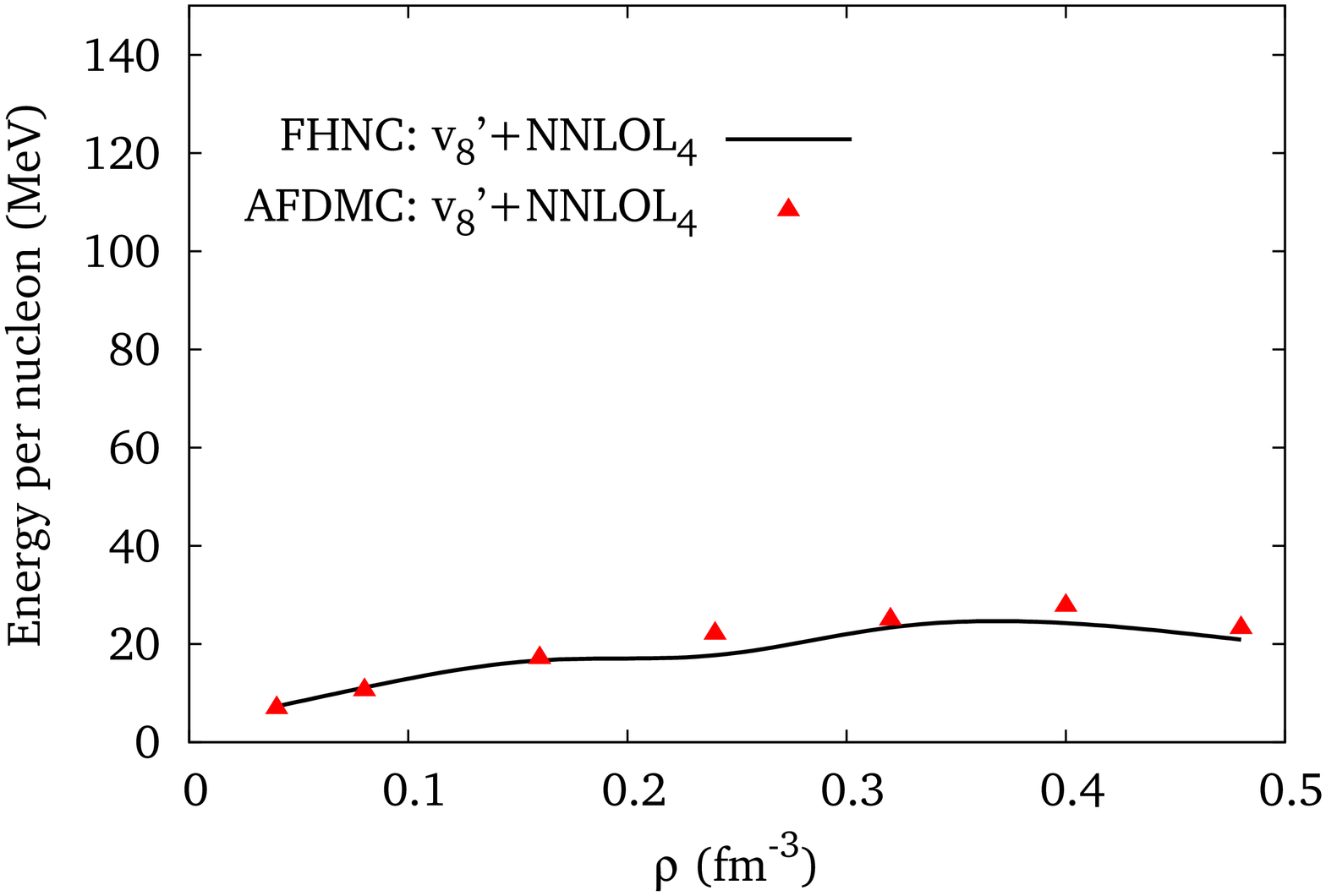}
\caption{Same as Fig. \ref{fig:tm_compare_PNM}, but for NNLOL plus $v_8^\prime$ hamiltonian.\label{fig:chiral_compare_PNM}}
\end{figure*}

The results displayed in Fig. \ref{fig:chiral_compare_PNM} show that, as in the case of the TM$^\prime$ potentials, the EoS of PNM computed within the AFDMC and FHNC/SOC schemes are very close to each other over the entire density range.

The EoS of Fig. \ref{fig:chiral_compare_PNM} are softer than those obtained from both the TM$^\prime$ (compare to Fig. \ref{fig:tm_compare_PNM}), 
and  $UIX$  (sse, e.g. Fig. 12 of  Ref.~\cite{lovato_11}) potentials. This is due to the ambiguity in the term $V_E$,  discussed in Section \ref{sec:cont_issue}. 

In the NNLOL$_2$, NNLOL$_3$, and NNLOL$_4$ models the constant $c_E$ is negative. Therefore, the contribution of $V_E$ is attractive, making the EoS very soft.  
When $V_E$ is repulsive (i.e. $c_E$ is positive), as in the NNLOL$_1$ potential, its contribution is very small and the resulting EoS, while being stiffer than those 
corresponding to the other NNLOL potentials, remains very soft.

The recent astrophysical data of Ref.~ \cite{demorest_10} suggest that the EoS of PNM be at least as stiff as the one obtained with a readjusted version of the effective density-dependent potential of Lagaris and Pandharipande in combination with the Argonne $v_{6}^\prime$ two-body interaction \cite{gandolfi_10}. Therefore, the EoS resulting from chiral NNLOL potentials are not likely to describe a neutron star of mass around $2M_\odot$.

\begin{figure}[!h]
\vspace{0.2cm}
\begin{center}
\includegraphics[angle=270,width=9.0cm]{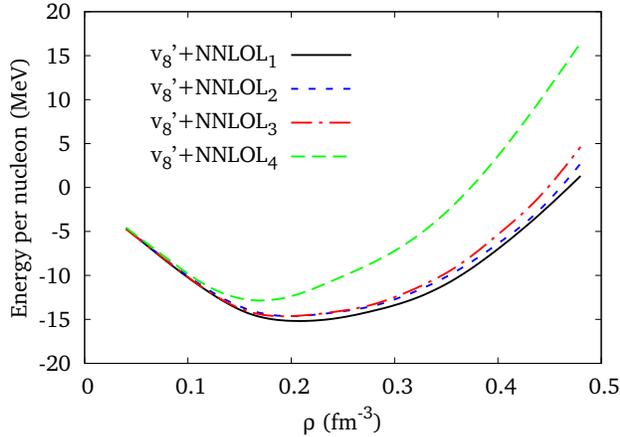}
\caption{Same as Fig. \ref{fig:tm_snm}, but for NNLOL plus $v_8^\prime$ hamiltonian.  \label{fig:nnlol_snm}}
\end{center}
\end{figure}

The SNM EoS corresponding to the NNLOL potentials are displayed in Fig. \ref{fig:nnlol_snm}. The fact that the NNLOL$_4$ potential provides the stiffest EoS, while in PNM provided the softest, is not surprising. As discussed in Section \ref{sec:cont_issue}, when the contact term is attractive in PNM, it is repulsive in SNM, and {\em viceversa}.

The results listed in Table \ref{tab:parameters_nnlol} show that none of the chiral NNLOL potentials fulfills the empirical constraints on the SNM EoS. 
All potentials overestimate the saturation density, while the compressibility is compatible with the empirical value only for the NNLOL$_2$ and NNLOL$_3$ models. 
As for the binding energies, they are closer to the experimental value than those obtained using both the UIX and TM$^\prime$ potentials. 

As a final remark, it has to be noticed that using the scalar repulsive term $V_{E}^I$ instead of $V_{E}^\tau$ provides more repulsion, resulting a stiffer EoS. 
As stressed in Section \ref{sec:cont_issue}, this issue needs to be addressed, taking into account all terms that become equivalent in the limit of infinite cutoff only. 
Moreover, since the discrepancies among these terms are of the same order as the NNNLO term of the chiral expansion, other contact terms have to be included  \cite{girlanda_11}.

\begin{table}[!h]
\caption{Saturation density, the binding energy per particle, and the compressibility related to the NNLOL Eos displayed in Fig. \ref{fig:nnlol_snm}. \label{tab:parameters_nnlol}}
\vspace{0.3cm}
\begin{tabular}{c c c c c} 
\hline 
 	& NNLOL$_1$ & NNLOL$_2$ & NNLOL$_3$ & NNLOL$_4$ \\
\hline
$\rho_0$ (fm$^{-3}$)&  0.21  &  0.20  &  0.19 & 0.17 \\

$E_0$ (MeV) &  -15.2  &  -14.6  &  -14.6 &  -12.9\\

K (MeV) &  198  &  252  &  220  & 310\\ 
\hline
 
\end{tabular} 
\vspace{0.1cm}
\end{table}

\section{Conclusions}
\label{conclusions}

A new generation of chiral inspired three-nucleon potentials in coordinate space, suitable for carrying out nuclear matter calculations, is now available. We have carried out a comparative analysis of the EoS of PNM and SNM obtained using the different parametrizations of the NNLOL potential, as well 
as the improved versions of the TM model discussed in Ref. \cite{kievsky_10}. 

The calculation of the SNM EoS has been been performed within the variational FHNC/SOC approach. In the case of PNM
we have also used the AFDMC computational scheme, the results of which turn out to be in close agreement with the variational FHNC/SOC estimates.

Our analysis shows that the transformation from momentum to coordinate space brings about a cutoff dependence, leading to sizable effects in 
nuclear matter. As discussed in Section \ref{sec:cont_issue}, the contribution of the contact term, which in PNM would vanish in the $\Lambda \to \infty$ limit, can not be fully determined fitting the low energy observables. Moreover, the NNN contact terms of the NNLOL$_2$ and NNLOL$_3$ models turn out to be attractive in PNM, leading to a strong softening of the EoS. 

An illustrative example of the uncertainty associated with the local form of the NNN contact term is provided by the results of Fig. \ref{fig:nnlol_snm} and 
Table \ref{tab:parameters_nnlol}. The NNLOL$_4$ model largely overestimates the empirical value of the compressibility modulus of SNM, thus yielding
a stiff EoS. On the other hand, as pointed out in Section \ref{sottosezione}, it predicts a soft EoS of PNM. 
The impact of this is ambiguity is large, since compressibility is a most important property of the EoS. The recent discovery of a $\sim 2$ M$_\odot$ 
neutron star appears in fact to rule out dynamical models yielding a soft  EoS of $\beta$-stable matter. 

None of the considered three-nucleon potential models simultaneously explains the empirical equilibrium density and binding energy of SNM. 
However, among the different parametrization that we have analyzed, the NNLOL$_4$ and TM$_3^\prime$ provide reasonable values of 
$\rho_0$. It has to be emphasized that this is a remarkable result, as, unlike the UIX model, 
these potential do not involve any parameter adjusted to reproduce $\rho_0$.  

In order to resolve  the inconsistencies involved in the contact term, one should include all contributions to this term arising from the chiral expansion at NNLO. 
Moreover, as pointed out by the authors of Ref.~\cite{girlanda_11}, due to the choice of the regulator function (see Eq.(\ref{eq:chi_cut})), a fully consistent treatment 
should also take into account NNNLO contact contributions. 

As a final remark, it must be mentioned that the  TM$^\prime$ and NNLOL potentials discussed in this paper can be used to 
obtain two-body density-dependent effective interaction within the formalism developed in Ref. \cite{lovato_11}. 
At present, this is the only approach allowing for the inclusion of three-nucleon potentials involving a term of the form of $V_4$ of Eq.(\ref{eq:2pi_3body_conf}) in AFDMC calculations of SNM. 


\vspace{1cm}
\begin{acknowledgements}
The authors are grateful to L. Asti, E. Epelbaum, L. Girlanda and A. Kievsky for many clarifying discussions. 
This work was partially funded under MIUR PRIN grant ``Many-body theory of nuclear systems and implications on the 
physics of neutron stars'',
 and National
Science Foundation grants PHY0757703 and PHY1067777.
The work of A.L. was partially supported by CompStar, a Research Networking Programme of the European Science
Foundation. 
\end{acknowledgements}


\begin{thebibliography}{99}
\bibitem{lacombe_80}M. Lacombe, B. Loiseau, J. M. Richard, R. Vinh Mau, J. C\^ont\'e , P. Pir\`es, and R. de Tourreil, Phys. Rev. C {\bf 21}, 861 (1980). 
\bibitem{stocks_94}V. G. J. Stoks, R. A. M. Klomp, C. P. F. Terheggen, and J. J. de Swart, Phys. Rev. C {\bf 49}, 2950 (1994).
\bibitem{wiringa_95}R.B. Wiringa, V.G.J. Stoks, R. Schiavilla, Phys. Rev. C {\bf 51}, 38 (1995).
\bibitem{machleidt_01}R. Machleidt, Phys. Rev. C {\bf 63}, 024001 (2001).
\bibitem{entem_03}D. R. Entem and R. Machleidt, Phys. Rev. C {\bf 68}, 041001(R) (2003).
\bibitem{epelbaum_05}E. Epelbaum, W. Gl\"ockle, Ulf-G. Mei\ss ner, Nucl. Phys. A {\bf 747}, 362 (2005).
\bibitem{stocks_93}V.G.J. Stoks, R.A.M. Klomp, M.C.M. Rentmeester, and J.J. de Swart, Phys. Rev. C {\bf 48},
792 (1993).
\bibitem{gabioud_79}B. Gabioud et al., Phys. Rev. Lett. {\bf 42}, 1508 (1979).
\bibitem{vanderleun_82}C. van der Leun and C. Alderlisten, Nucl. Phys. A {\bf 380}, 261 (1982).
\bibitem{ericson_83}T.E.O. Ericson and M. Rosa-Clot, Nucl. Phys. A {\bf 405}, 497 (1983).
\bibitem{rodning_90}N.L. Rodning and L.D. Knutson, Phys. Rev. C {\bf 41}, 898 (1990).
\bibitem{simon_81}G.G. Simon, Ch. Schmitt, and V.H. Walther, Nucl. Phys. A {\bf 364}, 285 (1981).
\bibitem{bishop_79}D.M. Bishop and L.M. Cheung, Phys. Rev. A {\bf 20}, 381 (1979).
\bibitem{shoen_03}K. Schoen et al., Phys. Rev. C {\bf 67}, 044005 (2003).
\bibitem{shimizu_95}S. Shimizu et al., Phys. Rev. C {\bf 52}, 1193 (1995).
\bibitem{epelbaum_02}E. Epelbaum, A. Nogga, W. Gl\"ockle, H. Kamada, Ulf-G. Mei\ss ner, and H. Vitala, Phys. Rev. C {\bf 66}, 064001 (2002).
\bibitem{bernard_08}V. Bernard, E. Epelbaum, H. Krebs, Ulf-G. Mei\ss ner, Phys. Rev. C {\bf 77}, 064004 (2008).
\bibitem{navratil_07}P. Navratil, Few Body Syst {\bf 41}, 117 (2007).
\bibitem{kievsky_10}A. Kievsky, M. Viviani, L. Girlanda, and L. E. Marcucci, Phys. Rev. C {\bf 81} 044003 (2010).
\bibitem{demorest_10}P. B. Demorest, T. Pennucci, S. M. Ransom, M. S. E. Roberts, and J. W. T. Hessels, Nature
{\bf 467}, 1081 (2010).
\bibitem{carlson_83}J. Carlson, V. R. Pandharipande, R. B. Wiringa, Nucl. Phys.{\bf A401}, 59 (1983).
\bibitem{coon_81}S. A. Coon and W. Gl\"ockle, Phys. Rev. C {\bf 23} 1790 (1981)
\bibitem{wiringa_88}R. B. Wiringa, V. Fiks, and A. Fabrocini, Phys. Rev. C {\bf 38}, 1010 (1988).
\bibitem{akmal_98}A. Akmal, V.R. Pandharipande, and D.G. Ravenhall, Phys. Rev. C {\bf 58}, 1804 (1998).
\bibitem{pudliner_95}B.S. Pudliner, V.R. Pandharipande, J. Carlson and R.B. Wiringa, Phys. Rev. Lett. {\bf 74}, 4396 (1995).
\bibitem{lovato_11}A. Lovato, O. Benhar, S. Fantoni, A. Yu. Illarionov, and K. E. Schmidt, Phys. Rev. C {\bf 83}, 054003 (2011).
\bibitem{gandolfi_07}S Gandolfi, F Pederiva, S Fantoni and K E Schmidt Phys. Rev. Lett. {\bf 98}, 102503 (2007).
\bibitem{bogner_05}S. K. Bogner, A. Schwenk, R. J. Furnstahl, A. Nogga, Nucl. Phys. A {\bf 763}, 59 (2005).
\bibitem{hebeler_11}K. Hebeler, S. K. Bogner, R. J. Furnstahl, A. Nogga, A. Schwenk, Phys. Rev. C {\bf 83}, 031301 (2011).
\bibitem{hebeler_10}K. Hebeler, A. Schwenk, Phys. Rev. C {\bf 82}, 014314 (2010).
\bibitem{weinberg_90} S. Weinberg, Phys. Lett. {\bf B251}, 288 (1990).
\bibitem{weinberg_91}S. Weinberg, Nucl. Phys. {\bf B363}, 3 (1991)
\bibitem{fettes_98}N. Fettes, U.-G. Mei\ss ner, S. Steiniger, Nucl. Phys. {\bf A640}, 199 (1998).
\bibitem{buttiker_00}P. B\"uttiker and U.-G. Mei\ss ner, Nucl. Phys. {\bf A668}, 97 (2000).
\bibitem{coon_01}S. A. Coon, H. K. Ha, Few Body Syst {\bf 30}, 131 (2001). 
\bibitem{pieper_01}S.C. Pieper, V.R. Pandharipande, R.B. Wiringa and J. Carlson, Phys. rev. C {\bf 64}, 014001 (2001). 
\bibitem{friar_99}J.L. Friar, D. H\"uber, and U. van Kolck, Phys. Rev. C {\bf 59}, 53 (1999).
\bibitem{fujita_57}J. Fujita and H. Miyazawa, Prog. Theor. Phys. {\bf 17}, 360 (1957).
\bibitem{wiringa_78}R. B. Wiringa and V. R. Pandharipande, Nucl. Phys. {\bf A299}, 1 (1978).
\bibitem{pandharipande_79}V. R. Pandharipande and R. B. Wiringa, Rev. Mod. Phys. {\bf 51}, 821 (1979).
\bibitem{schmidt_99}K. E. Schmidt and S. Fantoni, Phys. Lett. B446, 99 (1999).
\bibitem{pederiva_04}F. Pederiva, A. Sarsa, K. E. Schmidt and S. Fantoni, Nucl. Phys. A {\bf 742}, 255 (2004).
\bibitem{gandolfi_08}S. Fantoni, S. Gandolfi, A Yu. Illarionov, K. E. Schmidt and F. Pederiva, arxiv:0807.5043 .
\bibitem{gandolfi_09}S. Gandolfi, A. Yu. Illarionov, K. E. Schmidt, F. Pederiva and S. Fantoni, Phys. Rev. {\bf C79}, 054005 (2009).
\bibitem{gandolfi_10}S. Gandolfi, A. Yu. Illarionov, S. Fantoni, J. C. Miller, F. Pederiva, and K. E. Schmidt, MNRAS, Phys. Rev. {\bf 404}, L35 (2010).
\bibitem{girlanda_11}L. Girlanda, A. Kievsky, and M. Viviani, arXiv:1102.4799v2.


\end{thebibliography}
\end{document}